\documentclass[a4paper,11pt]{article}
\pdfoutput=1 

\usepackage{jcappub} 

\usepackage[T1]{fontenc} 
\title{Updated Bounds on Sum of Neutrino Masses in Various Cosmological Scenarios}

\author[a,b,1]{Shouvik Roy Choudhury,\note{Corresponding author.}}
\author[a,b,c]{Sandhya Choubey}

\affiliation[a]{Harish-Chandra Research Institute\\Chhatnag Road, Jhunsi, Allahabad 211019, India}
\affiliation[b]{Homi Bhabha National Institute\\ Training School Complex, Anushaktinagar, Mumbai - 400094, India}
\affiliation[c]{Department of Physics, School of Engineering Sciences,
KTH Royal Institute of Technology\\AlbaNova University Center, 106 91 Stockholm, Sweden}

\emailAdd{shouvikroychoudhury@hri.res.in}
\emailAdd{sandhya@hri.res.in}

\abstract{We present strong bounds on the sum of three active neutrino masses ($\sum m_{\nu}$) using selected cosmological datasets and priors in various cosmological models. We use the following baseline datasets: Cosmic Microwave Background (CMB) temperature data from Planck 2015, Baryon Acoustic Oscillations measurements from SDSS-III BOSS DR12, the newly released Type Ia supernovae (SNe Ia) dataset from Pantheon Sample, and a prior on the optical depth to reionization from 2016 Planck Intermediate results. We constrain cosmological parameters with these datasets with a Bayesian analysis in the background of $\Lambda CDM$ model with 3 massive active neutrinos. For this minimal $\Lambda CDM+\sum m_{\nu}$ model we find a upper bound of $\sum m_{\nu} <$ 0.152 eV at 95$\%$ C.L. Adding the high-$l$ polarization data from Planck strengthens this bound to  $\sum m_{\nu} <$ 0.118 eV, which is very close to the minimum required mass of $\sum m_{\nu} \simeq$ 0.1 eV for inverted hierarchy. This bound is reduced to $\sum m_{\nu} <$ 0.110 eV when we also vary r, the tensor to scalar ratio ($\Lambda CDM+r+\sum m_{\nu}$ model), and add an additional dataset, BK14, the latest data released from the Bicep-Keck collaboration (which we add only when $r$ is varied). This bound is further reduced to $\sum m_{\nu} <$ 0.101 eV in a cosmology with non-phantom dynamical dark energy ($w_0 w_a CDM+\sum m_{\nu}$ model with $w(z)\geq -1$ for all $z$). Considering the $w_0 w_a CDM+r+\sum m_{\nu}$ model and adding the BK14 data again, the bound can be even further reduced to $\sum m_{\nu} <$ 0.093 eV. For the $w_0 w_a CDM+\sum m_{\nu}$ model without any constraint on $w(z)$, the bounds however relax to $\sum m_{\nu} <$ 0.276 eV. Adding a prior on the Hubble constant ($H_0 = 73.24\pm 1.74$ km/sec/Mpc) from Hubble Space Telescope (HST), the above mentioned bounds further improve to $\sum m_{\nu} <$ 0.117 eV, 0.091 eV, 0.085 eV, 0.082 eV, 0.078 eV and 0.247 eV respectively. This substantial improvement is mostly driven by a more than 3$\sigma$ tension between Planck 2015 and HST measurements of $H_0$ and should be taken cautiously.}

\begin{document}
\maketitle
\flushbottom

\section{Introduction}
\label{sec:level1}
Neutrino oscillation experiments have put the existence of neutrino mass on a solid footing \cite{PhysRevLett.112.061802,PhysRevLett.108.191802,PhysRevD.86.052008,PhysRevLett.108.171803,PhysRevLett.94.081801,PhysRevLett.101.131802,PhysRevLett.81.1562,PhysRevLett.89.011301,CAPOZZI2016218}. There are three mass eigenstates ($\nu_1$, $\nu_2$, and $\nu_3$) which are quantum superpositions of the 3 flavour eigenstates ($\nu_e$, $\nu_{\mu}$, and $\nu_{\tau}$). The absolute neutrino mass scale, defined as the sum of the mass of the neutrino mass eigenstates, is the quantity,
\begin{equation}
\sum m_{\nu} \equiv m_1+ m_2+ m_3,
\end{equation}
where $m_i$ is the mass of the $i^{th}$ neutrino mass eigenstate. Tightest bounds on $\sum m_{\nu}$ come from cosmology.  

Massive but light active neutrinos behave as radiation in the early universe, when their mass is much higher than the temperature. Their energy density adds a contribution to the total radiation energy density ($\rho_r$) of the universe, which is conventionally parametrized by $N_{\textrm{eff}}$, an effective number of species of neutrinos,
\begin{equation}
\rho_r = \frac{\pi^2}{15}\left[ 1+ \frac{7}{8} \left(\frac{4}{11}\right)^{\frac{4}{3}} N_{\textrm{eff}} \right] T_{\gamma}^4,
\end{equation}
where $T_{\gamma}$ is the temperature of the photon. In the event of Neutrinos decoupling instantaneously from the QED plasma, we would have gotten $N_{\textrm{eff}} =$ 3. But neutrinos decouple in the early universe over many Hubble times (temperature T $\sim$ 10 MeV - 0.1 MeV), which changes $N_{\textrm{eff}}$ by a small amount, with the current best theoretical estimate being $N_{\textrm{eff}} =$ 3.046. This result is mostly due to (1) the continuation of decoupling of neutrinos during electron-positron annihilation and (2) QED plasma effects (see \cite{1610.02743,MANGANO2005221,PhysRevD.93.083522} for discussions on this topic.)
As long as neutrinos can be considered as a radiation species during the photon decoupling (temperature T$\sim$ 0.2 eV), CMB data shows no evidence of bounds on $N_{\textrm{eff}}$ not being compatible with its theoretically predicted value \cite{Planck2015}. Any departure of $N_{\textrm{eff}}$ from the theoretical prediction would be due to non-standard effects in the active neutrino sector or to the contribution of other relativistic species like a sterile neutrino. However in this work, we are only interested in bounds $\sum m_{\nu}$. We have not considered variation of $N_{\textrm{eff}}$. So in all our analyses we have fixed the value of $N_{\textrm{eff}}$ to 3.046.  

When neutrinos turn non-relativistic at late times, their energy density adds to the total matter density. Effect of massive neutrinos on cosmology has been widely studied in the literature \cite{LESGOURGUES2006307,doi:10.1146/annurev-nucl-102010-130252,Lesgourgues-Pastor,ABAZAJIAN201566,1367-2630-16-6-065002,1475-7516-2017-02-052}. For masses much smaller than 1 eV, neutrinos are still very much relativistic at the time of photon decoupling, and their mass cannot affect the evolution of photon perturbations.
Consequently, the effect can only appear on the background evolution, and secondary anisotropies like Integrated Sachs-Wolfe (ISW) Effect. These can be partially compensated by varying other free parameters of $\Lambda$CDM model, and hence CMB anisotropy alone is not a very useful tool to obtain strong bounds on $\sum m_{\nu}$. As they become non-relativistic, sub-eV neutrinos affect late time evolution of matter perturbations considerably. Due to neutrino free streaming effects, increasing suppression of matter power spectrum in small scales happens with increasing neutrino density fraction with respect to the matter density \cite{1367-2630-16-6-065002}. Therefore, CMB data accompanied with datasets from Baryon Acoustic Oscillations (BAO) and Large Scale Structure (LSS) measurements can produce very strong upper bounds on $\sum m_{\nu}$. 

Another neutrino property that future cosmological data might be able to determine is neutrino mass hierarchy. Different global fits \cite{PhysRevD.90.093006,Esteban2017,PhysRevD.95.096014,PhysRevD.96.073001,Gonzalez-Garcia2014} to neutrino oscillation experiment results have determined two mass squared splittings with considerable accuracy: $\Delta m_{21}^2 \equiv m_2^2 -m_1^2 \simeq 7.49^{+0.19}_{-0.17} \times 10^{-5}$ eV${^2}$ and $|\Delta m_{31}|^2 \equiv |m_3^2 -m_1^2| \simeq 2.484^{+0.045}_{-0.048} \times 10^{-3}$ eV${^2}$ (1$\sigma$ uncertainties) \cite{1475-7516-2016-11-035}. These are the mass squared splittings dictating the solar and atmospheric transitions respectively. Since the sign of $\Delta m_{31}^2$ is not known, it can be either +ve, giving rise to normal hierarchy (NH) with $m_1 < m_2 \ll m_3$, or -ve, giving rise to inverted hierarchy (IH) with $m_3 \ll m_1<m_2$. A lower bound can be put on $\sum m_{\nu}$ for each of the two scenarios. For NH, it is around $0.0585 \pm 0.00048$ eV, for IH it is around $0.0986 \pm 0.00085$ eV. These can be calculated from the values of the mass splittings by noting that
\begin{equation}
\sum m_{\nu} = m_0 + \sqrt{\Delta m_{21}^2+m_0^2} + \sqrt{|\Delta m_{31}|^2+m_0^2}~~~~~~~~~~~~~~~~~\textrm{\emph{(NH)}},
\end{equation}
and
\begin{equation}
\sum m_{\nu} = m_0 + \sqrt{|\Delta m_{31}|^2+m_0^2} + \sqrt{|\Delta m_{31}|^2 + \Delta m_{21}^2 + m_0^2} 
~~~\textrm{\emph{(IH)}},
\end{equation}
where $m_0$ is the lightest neutrino mass. By convention, $m_0 \equiv m_1$ for NH, and $m_0 \equiv m_3$ for IH. Substituting $m_0 = 0$ gives the minimum sum of masses for each case.

It is to be noted that current cosmological measurements are primarily sensitive to the sum of the three masses, $\sum m_{\nu}$. If the three active neutrinos have different masses, then they will each become non-relativistic at different temperatures, and will produce different levels of suppression to the small scale matter power spectrum. Therefore, same total mass, but different mass splittings should result in slightly different matter power spectra. However current experiments do not have the sensitivity to distinguish these differences with reasonable significance \cite{1475-7516-2016-11-035}. For this reason, in this work, we present results which are obtained with the approximation of 3 degenerate neutrino masses (from now on DEG), i.e.,
\begin{equation}
m_1 = m_2 = m_3 = \frac{\sum m_{\nu}}{3}~~~(DEG).
\end{equation}
This approximation is predominant in literature in analyses where $\sum m_{\nu}$ is varied. Planck data combined with others led to a bound of $\sum m_{\nu} <$ 0.23 eV at 95$\%$ C.L. (PlanckTT + lowP + lensing + BAO + JLA + H$_0$) \cite{Planck2015}. Depending on the used data and variations in the analysis, different analyses \cite{1475-7516-2015-11-011,CUESTA201677,Huang2016,PhysRevD.93.083527,PhysRevD.94.083522,PhysRevD.96.123503,Wang:2017htc} obtain 95$\%$ C.L. upper bounds from current data approaching the value of 0.1 eV, minimum mass required for IH. These results suggest IH is under pressure from cosmology. However, getting a 95$\%$ limit of $\sum m_{\nu}$ less than the minimum required mass for IH does not rule out IH. Recent papers \cite{1475-7516-2016-11-035,PhysRevD.96.123503} have suggested a rigorous but simple statistical method of computing the confidence level at which the hypothesis of IH can be rejected against NH using results from both cosmological and oscillations data. These recent analyses indicate that cosmology does slightly prefer NH compared to IH, but no statistically significant conclusion can be reached yet. In this paper, we, however, do not perform this statistical analysis and concentrate only on obtaining bounds on $\sum m_{\nu}$. 

In this work we provide bounds on $\sum m_{\nu}$ in the background of five different cosmological models: (1) $\Lambda CDM + \sum m_{\nu}$ (2) $\Lambda CDM+r+\sum m_{\nu}$, where $r$ is the tensor to scalar ratio, (3)  $w_0 w_a CDM + \sum m_{\nu}$, where we assume Chevallier-Polarski-Linder (CPL) parametrization for dynamical dark energy instead of a simple cosmological constant, (4) $w_0 w_a CDM+\sum m_{\nu}$ with $w(z)\geq -1$, i.e., we restrict the $w_0-w_a$ parameter space to exclude phantom dark energy (a recent study \cite{Vagnozzi:2018jhn} also has explored this model with datasets different from ours), and (5) $w_0 w_a CDM+ r +\sum m_{\nu}$ with $w(z)\geq -1$, a model extended with both tensors and non-phantom dynamic dark energy. 

We use combinations of the following recent datasets: (1) Cosmic Microwave Background (CMB) temperature, polarization and their cross-correlation data from Planck 2015; (2) Baryon Acoustic Oscillations measurements from SDSS-III BOSS DR12, MGS and 6dFGS; (3) the newly released Type Ia supernovae (SNe Ia) luminosity distance measurements from Pantheon Sample; (4) the latest data released from the BICEP/Keck Collaboration for the BB mode of the CMB spectrum; (5) and also local measurements of the Hubble parameter ($H_0$) from the Hubble Space Telescope; (6) the 2016 measurement of the optical depth to reionization ($\tau$) obtained from the analysis of the data from High Frequency Instrument of the Planck satellite; and (7) the galaxy cluster data from the observation of the Sunyaev-Zel'dovich (SZ) signature from thee 2500 square degree South Pole Telescope Sunyaev Zel'dovich (SPT-SZ) survey. We emphasize here that apart from Planck 2015 and the two redshift priors, the other datasets have not been studied widely in literature for obtaining bounds on $\sum m_{\nu}$ and we are the first to use combinations of these above mentioned datasets and priors to obtain the very strong bounds presented in this paper, in the above mentioned cosmological models.  

The two low redshift priors (on $\tau$ and $H_0$) are particularly important in constraining $\sum m_{\nu}$ because of presence of significant degeneracy of these two parameters with $\sum m_{\nu}$ \cite{PhysRevD.93.083527} in the CMB data. Also to be noted that, future  high-resolution  CMB  polarization  measurements \cite{Matsumura2014,1475-7516-2011-07-025,1475-7516-2018-04-015,1610.02743} might be able to reconstruct the lensing potential very accurately and provide even stronger constraints on $\sum m_{\nu}$. However, we do not include the Planck lensing data, as the lensing potential measurements via reconstruction through the four-point functions of CMB data from Planck 2015 are in tension with the constraints obtained from CMB power spectra \cite{Planck2015}. 

This paper is structured as follows: in Section~\ref{sec:level2} we describe our analysis method, the varying parameters of various cosmological models analyzed in this paper and the priors on the said parameters. We also briefly describe the Chevallier-Polarski-Linder (CPL) parametrization for dynamical dark energy. In Section~\ref{sec:level3}, we briefly describe the various datasets we have used in this work. In Section~\ref{sec:level4} we present the results of our analysis. We conclude in Section~\ref{sec:level5}. 

\section{Cosmological Models and Analysis Method}
\label{sec:level2} 
As mentioned in the previous section, in this work we have considered 5 different models of cosmology to obtain bounds on the sum of three active neutrino masses. Below we list the vector of parameters to vary in each of these cosmological models. 

For $\Lambda CDM + \sum m_{\nu}$ model: 
\begin{equation}\label{eqn:1}
\theta \equiv \left[\omega_c, ~\omega_b, ~\Theta_s,~\tau, ~n_s, ~ln [ 10^{10} A_s], \sum m_{\nu} \right].
\end{equation} 

For $\Lambda CDM+r+\sum m_{\nu}$ model:
\begin{equation}\label{eqn:2}
\theta \equiv \left[\omega_c, ~\omega_b, ~\Theta_s,~\tau, ~n_s, ~ln [ 10^{10} A_s], \sum m_{\nu}, r \right].
\end{equation}

For both the $w_0 w_a CDM + \sum m_{\nu}$ models (with or without phantom dark energy) :
\begin{equation}\label{eqn:3}
\theta \equiv \left[\omega_c, ~\omega_b, ~\Theta_s,~\tau, ~n_s, ~ln [ 10^{10} A_s], \sum m_{\nu}, w_0, w_a \right].
\end{equation}

For the $w_0 w_a CDM + r + \sum m_{\nu}$ model (non-phantom dark energy with tensors) :
\begin{equation}
\theta \equiv \left[\omega_c, ~\omega_b, ~\Theta_s,~\tau, ~n_s, ~ln [ 10^{10} A_s], \sum m_{\nu}, w_0, w_a, r \right].
\end{equation}

Here the first 7 parameters for all the models are same. Out of them, the first 6 correspond to the $\Lambda CDM$ model. Here $\omega_c = \Omega_c h^2$  and  $\omega_b = \Omega_b h^2$ are the present-day physical CDM and baryon densities respectively. $\Theta_s$ is the  the ratio between the sound horizon
and the angular diameter distance at decoupling. $\tau$ is the optical depth to reionization. $n_s$ and $A_s$ are the power-law spectral index and power of the primordial curvature perturbations, respectively, at the pivot scale of $k_* = 0.05 h Mpc^{-1} $. The 7th parameter is of course $\sum m_{\nu}$ which is of our biggest concern in this work. 

In $\Lambda CDM+r+\sum m_{\nu}$, along with scalar perturbations we also include tensor perturbations and let another parameter $r$ to vary, which is the tensor-to-scalar ratio at the pivot scale of $k_* = 0.05 h Mpc^{-1} $. The choice to study this model is motivated by the results from the latest publicly available dataset from the measurement of the BB mode spectrum of the CMB from Bicep-Keck collaboration, namely BK14, which provides an upper bound to $r<$ 0.07 at 95\% C.L, when combined with Planck 2015 and other datasets \cite{PhysRevLett.116.031302}; while the bound without BK14 data is much more relaxed at $r<$ 0.12 \cite{Planck2015}. We expect this data to modify the constraints on the neutrino related parameters.

For the two dynamical dark dark energy models, we again only concentrate on scalar perturbations, but the background $\Lambda CDM$ cosmology with the dark energy equation of state (EoS) $w = -1$ is replaced by a varying equation of state with the following parametrization in terms of the redshift $z$:
\begin{equation}
w (z) = w_0 + w_a \frac{z}{1+z}.
\end{equation}
This parametrization is famously known as the  Chevallier-Polarski-Linder (CPL) parametrization \cite{doi:10.1142/S0218271801000822,PhysRevLett.90.091301}. This just uses the first two terms in a Taylor expansion of the EoS in powers of the scale factor $a = 1/(1+z)$. This parametrization is suitable for describing the past expansion history of the universe, especially at high redshifts \cite{PhysRevLett.90.091301}, but other parametrizations might be needed to describe future evolution \cite{MA2011233} since as $z \rightarrow -1$,  $w(z)$ diverges.

Notice that $w(z=0) = w_0$ corresponds to the dark energy EoS today, whereas $w(z\rightarrow \infty) = w_0+w_a$ refers to the dark energy EoS in the very far past.  Between these two times, it is easy to see that $w(z)$ is a monotonic function. Therefore, to explore only the non-phantom dark energy region of the parameter space, i.e., $w(z) \geq -1$, it is sufficient to apply the following hard priors \cite{Vagnozzi:2018jhn}:
\begin{equation}\label{eqn:4}
w_0 \geq -1;~~~~~~~w_0+w_a\geq -1.
\end{equation}
We abbreviate the $w_0 w_a CDM+\sum m_{\nu}$ model with $w(z)\geq -1$ for all $z$ as the NPDDE model, whereas $w_0 w_a CDM+\sum m_{\nu}$ model without any such restriction on the EoS will be simply called the DDE model. A Universe dominated by a phantom dark energy component ($w(z)\leq -1$) would end in a Big Rip in most cosmological models, where dark energy density becomes infinite in a finite time, resulting in dissociation of any bound state, i.e., the "Big Rip" \cite{Caldwell:2003vq}. Such a universe is unphysical in a sense and hence we study the NPDDE model separately. 

In our work, we conduct a Bayesian analysis to derive constraints on $\sum m_{\nu}$.  For all the parameters listed in Eq.~(\ref{eqn:1}), Eq.~(\ref{eqn:2}), and Eq.~(\ref{eqn:3}), we impose flat priors in our analysis. The prior ranges are listed on the  Table~\ref{table:1}. We obtain the posteriors using the Markov Chain Monte Carlo (MCMC) sampler CosmoMC \cite{PhysRevD.66.103511} which uses CAMB \cite{Lewis:1999bs} as the Boltzmann solver and the Gelman and Rubin statistics \cite{doi:10.1080/10618600.1998.10474787} to quantify the convergence of chains.

\begin{table}[tbp]
\centering
\begin{tabular}{|lc|} 
\hline
Parameter & Prior\\
\hline
$\omega_c$ & [0.001,0.99] \\
$\omega_b$ & [0.005,0.1]  \\
$\Theta_s$ & [0.5,10] \\
$\tau$ & [0.01,0.8]\\
$n_s$ & [0.8,1.2] \\
ln $[10^{10} A_s]$ & [2,4] \\
$\sum m_{\nu}$ (eV) & [0,5] \\
$r$ & [0,1]\\
$w_0$ & [-3, -0.33]\\
$w_a$ & [-2, 2]\\
\hline

\end{tabular}
\caption{\label{table:1} Flat priors on cosmological parameters included in this work. For the NPDDE model, hard priors according to Eq.~(\ref{eqn:4}) are also implemented so as to exclude the parameter space region corresponding to phantom dark energy.}
\end{table}

\section{Datasets}
\label{sec:level3} 
Below, we provide a description of the datasets used in our analyses. We have used different combinations of these datasets. 

\emph{Cosmic Microwave Background: Planck 2015}:

Measurements of the CMB temperature, polarization, and temperature-polarization cross-correlation spectra from the publicly available Planck 2015 data release \cite{Planck2015-I} are used. We consider a combination of the high-$l$ (30 $\leq$ $l$ $\leq$ 2508) TT likelihood, as well as the low-$l$ (2 $\leq$ $l$ $\leq$ 29) TT likelihood. We call this combination simply as TT. Along with that, we include the Planck polarization data in the low-$l$ (2 $\leq$ $l$ $\leq$ 29) likelihood, and refer to this as lowP. We also consider the high-$l$ (30 $\leq$ $l$ $\leq$ 1996) EE and TE likelihood. This dataset and TT together are referred to as TTTEEE. The high-$l$ polarization data might still be contaminated with residual systematics \cite{Planck2015}, so bounds obtained without the use of high-$l$ polarization can be considered slightly more reliable.

\emph{Baryon Acoustic Oscillations (BAO) Measurements and Related Galaxy Cluster data}:

In this work, we include BAO measurements obtained from various galaxy surveys. We make use of the SDSS-III BOSS DR12 Consensus sample (as described in \cite{doi:10.1093/mnras/stx721}; uses LOWZ and CMASS galaxy samples at $z_{\textrm{eff}} =$ 0.38, 0.51 and 0.61), the DR7 Main Galaxy Sample (MGS) at $z_{\textrm{eff}} = 0.15$ \cite{doi:10.1093/mnras/stv154}, and the Six-degree-Field Galaxy Survey (6dFGS) survey at $z_{\textrm{eff}} = 0.106$ \cite{doi:10.1111/j.1365-2966.2011.19250.x}. We refer to this combination as BAO. Here $z_{\textrm{eff}}$ is the effective redshift of the particular survey. In some cases, we have also used the full shape measurements of the correlation function and galaxy power spectrum (refer to \cite{doi:10.1093/mnras/stx721} for details) from the SDSS-III BOSS DR12. We denote this as FS. The full shape of these measurements reveal additional information other than the BAO signal. 

\emph{Type Ia Supernovae (SNe Ia) Luminosity Distance Measurements}: 

We also include Supernovae Type-Ia (SNe Ia) luminosity distance measurements from the Pantheon Sample \cite{Scolnic:2018rjj} which consists of data from  279 Pan-STARRS1 (PS1) Medium Deep Survey SNe Ia ($0.03 < z < 0.68$) and combines it with distance estimates of SNe Ia from SDSS, SNLS, various low-z and HST samples. This combined sample of SNe Ia is largest till date and consists of data from a total of 1048 SNe Ia with redshift range $0.01 < z < 2.3$. We denote this data set as PAN hereafter. This dataset replaces the Joint Light-curve Analysis (JLA) SNe Ia sample which consists of 740 spectroscopically confirmed type Ia supernovae \cite{JLA}. 

\emph{Galaxy Cluster Data from South Pole Telescope}: 

In this work, we use data from the SPT-SZ survey \cite{0004-637X-782-2-74} which provides data from a sample of 377 clusters (identified at $z > 0.25$). SPT-SZ is a survey of 2500 deg$^2$ of the southern sky conducted with the South Pole Telescope (SPT, \cite{1538-3873-123-903-568}). These galaxy clusters are recognized by their Sunyaev-Zel'dovich (SZ) effect \cite{BIRKINSHAW199997} signature. We call this dataset as SZ from now on.

\emph{Optical Depth to Reionization}:

The optical depth is proportional to the electron number density integrated along the line of sight, and thus most of the contribution to it comes from the time when the universe re-ionizes. We impose a bound on this optical depth to reionization, $\tau = 0.055 \pm 0.009$, taken from \cite{Planck2016-intermediate}, in which Planck collaboration has identified, modelled and removed previously unexplained systematic effects in the polarization data of the Planck High Frequency Instrument (HFI) on large angular scales (low-$l$) (the data was not made publicly available). It is currently the most recent and reliable measurement of $\tau$ from Planck data. We shall hereafter refer to this prior as $\tau 0p055$. We use $\tau 0p055$ as a substitute for low-$l$ polarization data, and hence we exclude the lowP data whenever we apply the $\tau 0p055$ prior, to avoid any double counting. 

\emph{Hubble Parameter Measurements}: 

We use a Gaussian prior of $73.24 \pm 1.74$ km/sec/Mpc  on $H_0$ from the recent 2.4\% determination of the local value of the Hubble constant by \cite{0004-637X-826-1-56}, combining the anchor NGC 4258, Milky Way and LMC Cepheids. We shall refer to this prior as R16. It is to be noted that different datasets prefer different values of $H_0$ and there is no clear consensus. For instance  recent strong lensing observations \cite{doi:10.1093/mnras/stw3006} of the H0LiCOW program give a slightly lower value of $H_0 = 71.9^{+2.4}_{-3.0}$ km/sec/Mpc, whereas another measurement \cite{0004-637X-835-1-86} prefers a much lower value of $H_0 = 68.3^{+2.7}_{-2.6}$ km/sec/Mpc. The recent SDSS DR12 BAO data prefers an even lower value of $67.6 \pm 0.5$ km/sec/Mpc \cite{doi:10.1093/mnras/stx721}. We chose the R16 value as it is in 3.4$\sigma$ tension with Planck 2016 intermediate results \cite{Planck2016-intermediate}, whose measured value of $H_0$ is $66.93 \pm 0.62$ km/sec/Mpc assuming $\Lambda CDM$ with 3 active neutrinos of total mass fixed at $\sum m_{\nu} = 0.06$ eV.  Using the R16 prior we get an idea of how the parameter bounds will change if cosmology has to accommodate such a large value of Hubble constant, say, through some new physics. 

\emph{B Mode Polarization data of CMB}: 

For the BB mode spectrum of CMB, we use the latest dataset available from BICEP/Keck collaboration which includes all data (spanning the range: $20 < l < 330$) taken up to and including 2014 \cite{PhysRevLett.116.031302}. This dataset is referred to as BK14. 

\section{Results on $\sum m_{\nu}$}
\label{sec:level4} 

For clarity, we have presented and explained the results on the sum of three active neutrino masses separately for each model (see Section~\ref{sec:level2} for a description of models) in different subsections. All the quoted upper bounds are at 95\% C.L. The main results are summarized in Tables~\ref{table:2} -- \ref{table:11}. Details about models and datasets are given in Section~\ref{sec:level2} and Section~\ref{sec:level3} respectively.

\subsection{Results for the $\Lambda CDM + \sum m_{\nu}$ Model }
\label{sec:level4:1} 
In this subsection, we present the 2$\sigma$ (95\% C.L.) upper bounds on $\sum m_{\nu}$ for the $\Lambda CDM + \sum m_{\nu}$ model for various combinations of datasets.  Upper bounds on $\sum m_{\nu}$ are given at 2$\sigma$ (95\% C.L.) while marginalized limits for any other parameter mentioned in the text are given at 1$\sigma$ (68\% C.L.). We have divided these results in two separate sections for convenience of analyzing and presenting. First in Section~\ref{sec:level4:1:1} we present results obtained without using any priors on the optical depth to reionization ($\tau$) and Hubble constant ($H_0$) and discuss the effects of different datasets on the bounds. In Section~\ref{sec:level4:1:2} we summarize the results obtained using the said priors.

\subsubsection{Results without $\tau$ and $H_0$ priors}
\label{sec:level4:1:1} 
\begin{figure}[tbp]
\centering 
\includegraphics[width=.4963\linewidth]{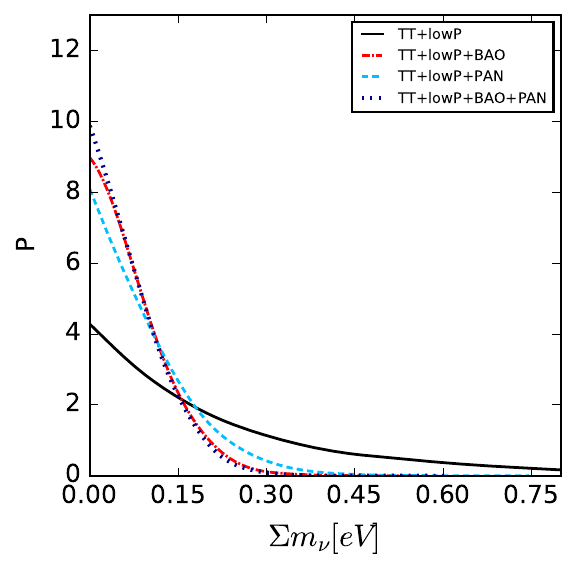}
\hfill
\includegraphics[width=.4963\linewidth]{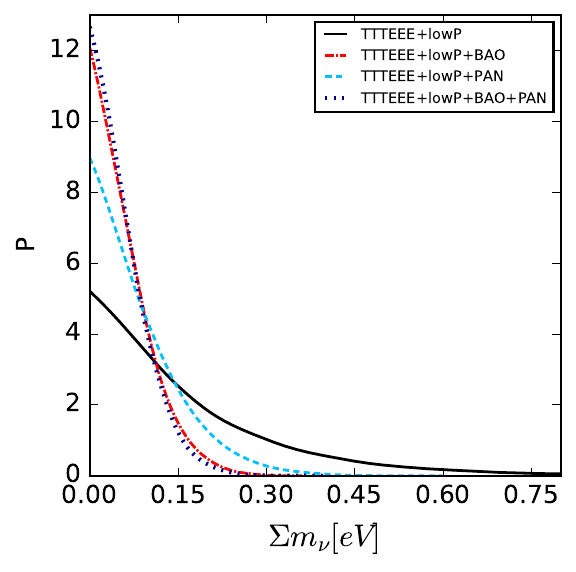}
\caption{\label{fig:1} Comparison of 1-D marginalized posterior distributions for $\sum m_{\nu}$ for various data combinations in $\Lambda CDM + \sum m_{\nu}$, without $\tau$ and $H_0$ priors. The plots are normalized in the sense that area under the curve is same for all curves. }
\end{figure}

\begin{figure}[tbp]
\centering 
\includegraphics[width=.4963\linewidth]{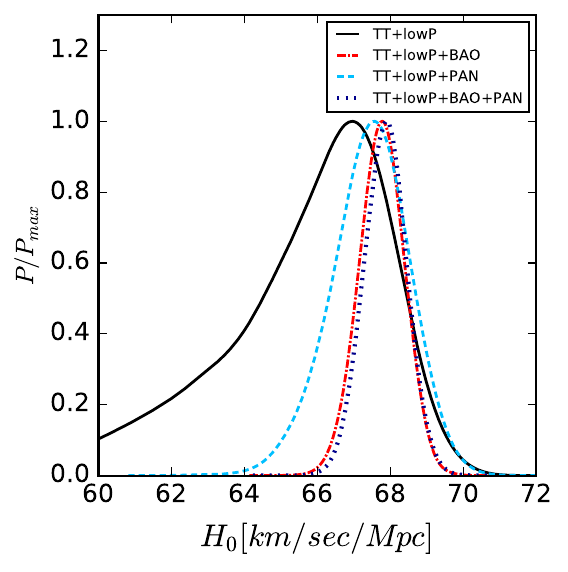}
\hfill
\includegraphics[width=.4963\linewidth]{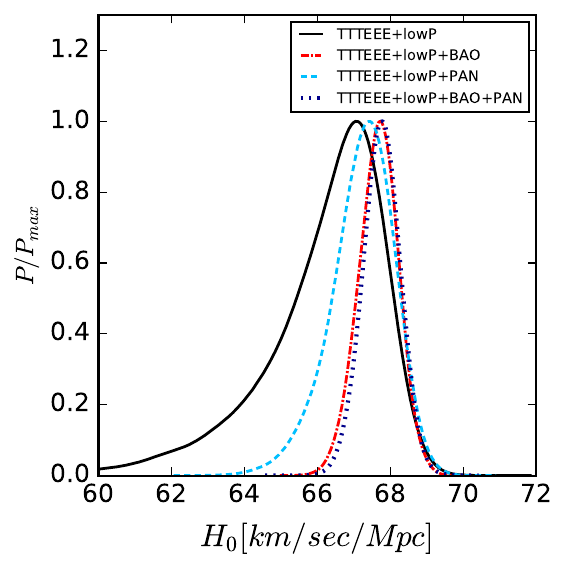}
\caption{\label{fig:2} Comparison of 1-D marginalized posterior distributions for $H_0$ for various data combinations in $\Lambda CDM + \sum m_{\nu}$, without $\tau$ and $H_0$ priors.}
\end{figure}

\begin{table}[]
\centering
\begin{tabular}{@{}|l|r|@{}}
\hline
\multicolumn{2}{|c|}{Model: $\Lambda CDM + \sum m_{\nu}$} \\ 
\hline
Dataset                 & $\sum m_{\nu}$ (95\% C.L.)    \\ 
\hline
TT                  & < 1.064 eV                      \\
TT + lowP           & < 0.724 eV                      \\
TT + BAO            & < 0.311 eV                              \\
TT + lowP + BAO     & < 0.200 eV                              \\
TT + PAN            & < 0.383 eV                              \\
TT + lowP + PAN     & < 0.260 eV                             \\
TT + BAO + PAN      & < 0.299 eV                             \\
TT + lowP + BAO + PAN      & < 0.190 eV                             \\
\hline
\end{tabular}
\caption{\label{table:2} 95\% C.L. upper bounds on sum of three active neutrino masses in the degenerate case, in the backdrop of $\Lambda CDM + \sum m_{\nu}$ model for the given datasets. Details about models and datasets are given in Section \ref{sec:level2} and Section~\ref{sec:level3} respectively. }
\end{table}

\begin{table}[tbp]
\centering
\begin{tabular}{@{}|l|r|@{}}
\hline
\multicolumn{2}{|c|}{Model: $\Lambda CDM + \sum m_{\nu}$} \\ 
\hline
Dataset                 & $\sum m_{\nu}$ (95\% C.L.)    \\ 
\hline
TTTEEE                  & < 0.833 eV                      \\
TTTEEE + lowP           & < 0.508 eV                      \\
TTTEEE + BAO            & < 0.204 eV                              \\
TTTEEE + lowP + BAO     & < 0.158 eV                              \\
TTTEEE + PAN            & < 0.306 eV                              \\
TTTEEE + lowP + PAN     & < 0.230 eV                              \\
TTTEEE + BAO + PAN      & < 0.196 eV                              \\
TTTEEE + lowP + BAO + PAN  & < 0.145 eV                            \\
\hline
\end{tabular}
\caption{\label{table:3} 95\% C.L. upper bounds on sum of three active neutrino masses in the degenerate case, in the backdrop of $\Lambda CDM + \sum m_{\nu}$ model for the given datasets.  This is same as Table~\ref{table:2} but including the high-$l$ polarization data of Planck 2015. Details about models and datasets are given in Sections~\ref{sec:level2} and \ref{sec:level3} respectively.}
\end{table}

In Tables~\ref{table:2} and \ref{table:3} we present the bounds without applying any Gaussian prior to the low redshift parameters $\tau$ and $H_0$. In Table~\ref{table:2} bounds are obtained without the use of the high-$l$ polarization data from Planck 2015, while in Table~\ref{table:3} it is included. Figure~\ref{fig:1} and \ref{fig:2} shows 1-D marginalized posterior distributions for $\sum m_{\nu}$ and $H_0$ respectively, for various data combinations. As mentioned in Section~\ref{sec:level1}, CMB TT data alone is not particularly sensitive to masses much lower than 1 eV. This is clearly reflected in the results. The TT data alone can only constrain $\sum m_{\nu} < 1.064$ eV at 95\% C.L.. Addition of the high-$l$ E mode polarization auto-correlation and temperature-polarization cross-correlation data leads to a higher constraining capability, reducing the bound to $\sum m_{\nu} < 0.833$ eV. This phenomenon of mass bounds getting stronger with addition of high-$l$ polarization data is seen throughout all the analyses we have done, and corroborates well with previous studies \cite{Planck2015,PhysRevD.96.123503}. 

\begin{figure}[tbp]
\centering 
\includegraphics[width=.6\linewidth]{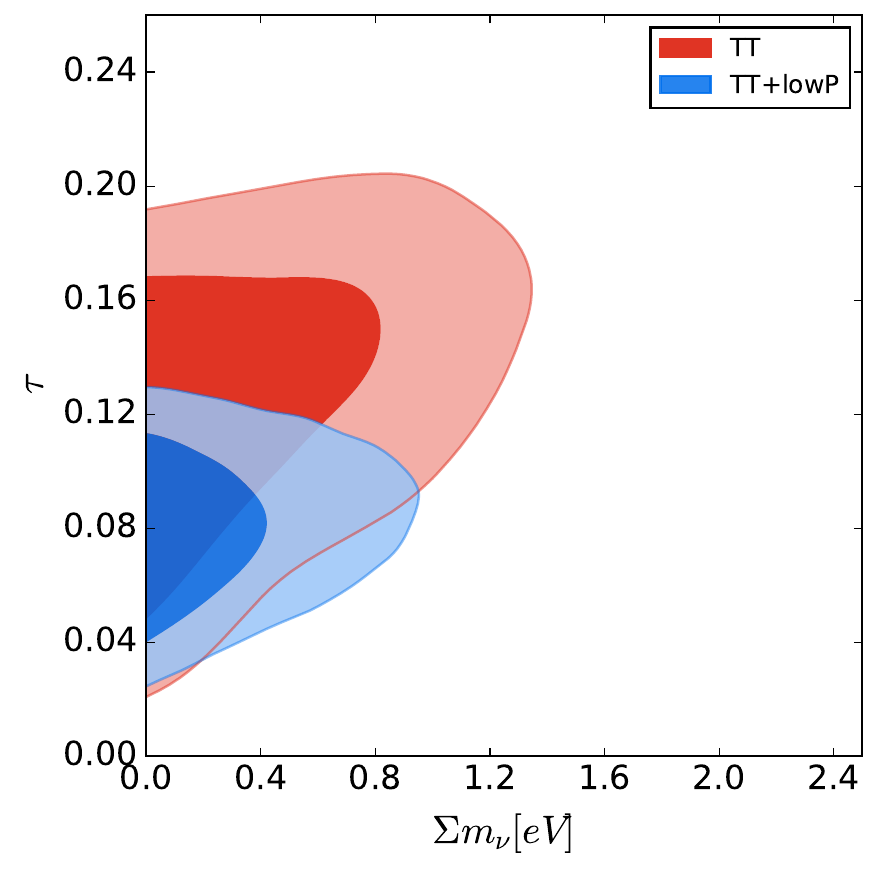}
\caption{\label{fig:3} 1$\sigma$ and 2$\sigma$ marginalised  contours for $\tau$ vs. $\sum m_{\nu}$ for TT and TT+lowP datasets in the $\Lambda CDM + \sum m_{\nu}$ model, showing the reduction in correlation between $\tau$ and $\sum m_{\nu}$ due to addition of lowP data, leading to a stronger bound on $\sum m_{\nu}$.}
\end{figure}

Addition of the lowP data makes the bounds significantly stronger, i.e., $\sum m_{\nu} < 0.724$ eV for TT+lowP and $\sum m_{\nu} < 0.508$  eV for TTTEEE+lowP. This can be attributed to lowP data being able to partially do away with degeneracies present with $\sum m_{\nu}$ and other parameters like $\tau$ and $A_s$. If we consider TT data only, an increase in $\sum m_{\nu}$ reduces the smearing of the damping tail \cite{Allison:2015qca,LEWIS20061}, which can be compensated by an increase in $\tau$. The value of $A_s$ also needs to increase, as the Planck TT data severely constrains the quantity $A_s e^{-2\tau}$, which leads to a degeneracy between these two parameters; variations approximately following the relation $\delta A_s/A_s\sim 2 ~\delta \tau$. Effects of $A_s$ and $\sum m_{\nu}$ are also not independent in cosmology. The value of $A_s$ determines the overall amplitude of matter power spectrum. Increase in $A_s$ increases the amplitude, whereas an increase $\sum m_{\nu}$ suppresses matter power spectrum in small scales. The low-$l$ polarization data can in principle break this degeneracy between $A_s$ and $\tau$, and consequently the three-way degeneracy between $A_s$, $\tau$ and $\sum m_{\nu}$. This is possible through the appearance of the well known "reionization bump" in the $l<20$ range in the polarization spectra whose amplitude is $\propto \tau^2$ in the EE spectra and $\propto \tau$ in the TE spectra \cite{Reichardt2016}, and the bump cannot be reproduced by varying other parameters, thus breaking the degeneracy. Indeed, while the TT data alone prefers a $\tau = 0.127^{+0.037}_{-0.033}$, the TT+lowP data prefers a much lower $\tau = 0.080 \pm 0.019$; a smaller value of $\tau$ leading to a stronger upper bound of $\sum m_{\nu}$. Refer to Figure~\ref{fig:3} for a visualization of this effect.  Similar inference can be made for TTTEEE and TTTEEE+lowP. However this degeneracy breaking is only partial. A very precise measurement of low-$l$ polarization is needed to completely break the degeneracy. 

While $\sum m_{\nu}$ and $\tau$ are strongly correlated in the Planck TT, $\sum m_{\nu}$ and $H_0$ are strongly anti-correlated. 
Defining $\omega_i \equiv \Omega_i h^2$ (where $i \equiv \gamma,c,b,\Lambda$ with $\gamma \equiv$ photons, $c \equiv$ CDM, $b \equiv$ baryons, and $\Lambda \equiv$ cosmological constant) the comoving distance to the last scattering surface at redshift $z_{dec}$ in a flat $\Lambda CDM+ \sum m_{\nu}$ universe is given by,
\begin{equation}
\label{eq:4.1}
\chi (z_{dec}) = \int^{z_{dec}}_0 \frac{dz}{H(z)} \propto \int^{z_{dec}}_0 \frac{dz}{\sqrt{\omega_{\gamma} (1+z)^4+ (\omega_c+\omega_b) (1+z)^3 + \omega_{\Lambda}+\frac{\rho_{\nu}(z)h^2}{\rho_{cr,0}}}},
\end{equation} 
where $\rho_{\nu}(z)$ is the neutrino energy density at a redshift $z$, and $\rho_{cr,0} = 3H_0^2/8\pi G$ is the critical density today. $\rho_{\nu}(z)$ scales differently with redshift, depending on whether neutrinos can be considered as radiation or matter. At late times, when neutrinos become non-relativistic, $\rho_{\nu}(z)$ scales as matter (i.e. $\rho_{\nu}(z) \propto (1+z)^3$)  and depends on $\sum m_{\nu}$. Since in a flat universe, $\Omega_{\Lambda} = 1 - (\Omega_c + \Omega_b)-\Omega_{\gamma} - \Omega_{\nu}$, at late times, the last two terms within the square root in the denominator in Eq.~\ref{eq:4.1} give: 
\begin{equation}
\omega_{\Lambda}+\frac{\rho_{\nu}(z)h^2}{\rho_{cr,0}} = (1 - \Omega_{\gamma})h^2 - (\omega_c + \omega_b) + \omega_{\nu} ((1+z)^3-1). 
\end{equation}
Now, $(\omega_c + \omega_b)$ is well constrained by CMB accoustic peaks. Since $\omega_{\nu} \propto \sum m_{\nu}$, any change to $\chi (z_{dec})$ due to increase in $\sum m_{\nu}$ can be compensated by decreasing $h$, i.e., $H_0$, and hence the anti-correlation.                                                                                                                                                                                                                                                                                                                                                  

Addition of the BAO data improves the mass bounds significantly by partially breaking the degeneracy between $\sum m_{\nu}$ and $H_0$. We find that addition of the BAO data to TT + lowP reduces the bound to $\sum m_{\nu} < 0.200$ eV from  $\sum m_{\nu} < 0.724$ eV. For the TTTEEE+lowP+BAO case, we get $\sum m_{\nu} < 0.158$ eV, which is also much stronger than the bound without BAO data. One can understand such important changes in bounds by understanding the impact of neutrino masses on the quantity $D_{\nu}(z_{\textrm{eff}})/r_s(z_{\textrm{drag}})$ which is measured by BAO using spatial correlation of galaxies. Here $r_s(z_{\textrm{drag}})$ is the comoving sound horizon at the end of the baryon drag epoch (the epoch at which baryons decouple from photons, slightly after recombination) and changes in $\sum m_{\nu}$ has a small effect on it. On the other hand, the dilation scale $D_{\nu}(z_{\textrm{eff}})$ at the effective redshift $z_{\textrm{eff}}$ of the survey, is a combination of the angular diameter distance $D_{A}(z)$ and the Hubble parameter $H(z)$,
\begin{equation}
D_{\nu}(z) = \left[ (1+z)^2 D_A^2(z) \frac{cz}{H(z)}\right]^{1/3}~~~~~~(c \equiv \textrm{speed of light}),
\end{equation}  
and it is the quantity which is affected by $\sum m_{\nu}$ most. If $\sum m_{\nu}$ is increased while $\omega_c+ \omega_b$ is kept fixed, the expansion rate at early times increases. This requires $\Omega_{\Lambda}$ to decrease to keep $\Theta_s$ fixed, which is very well constrained by the CMB power spectra. Decrease in $\Omega_{\Lambda}$ leads to a increase in $D_{\nu}(z_{\textrm{eff}})$, which in turn leads to a decrease in both $r_s(z_{\textrm{drag}})/D_{\nu}(z_{\textrm{eff}}))$ and $H_0$. BAO data prefers a higher value of $H_0$ than the CMB spectra, and by rejecting the lower $H_0$ values removes the regions with higher $\sum m_{\nu}$ values. See \cite{0004-637X-782-2-74} for a detailed discussion on this topic. In our analysis we found that TT+lowP data prefers a value of $H_0 = 65.53^{+3.01}_{-1.26}$ km/sec/Mpc, whereas TT+lowP+BAO prefers $H_0 = 67.76\pm 0.62$ km/sec/Mpc, confirming the above. For TTTEEE+lowP and TTTEEE+lowP+BAO these bounds are $H_0 = 66.17^{+1.96}_{-0.81}$ km/sec/Mpc and $H_0 = 67.67^{+0.54}_{-0.51}$ km/sec/Mpc respectively. This effect of BAO data rejecting lower $H_0$ values is evident from Figure~\ref{fig:2}.  

\begin{figure}[tbp]
\centering 
\includegraphics[width=.4963\linewidth]{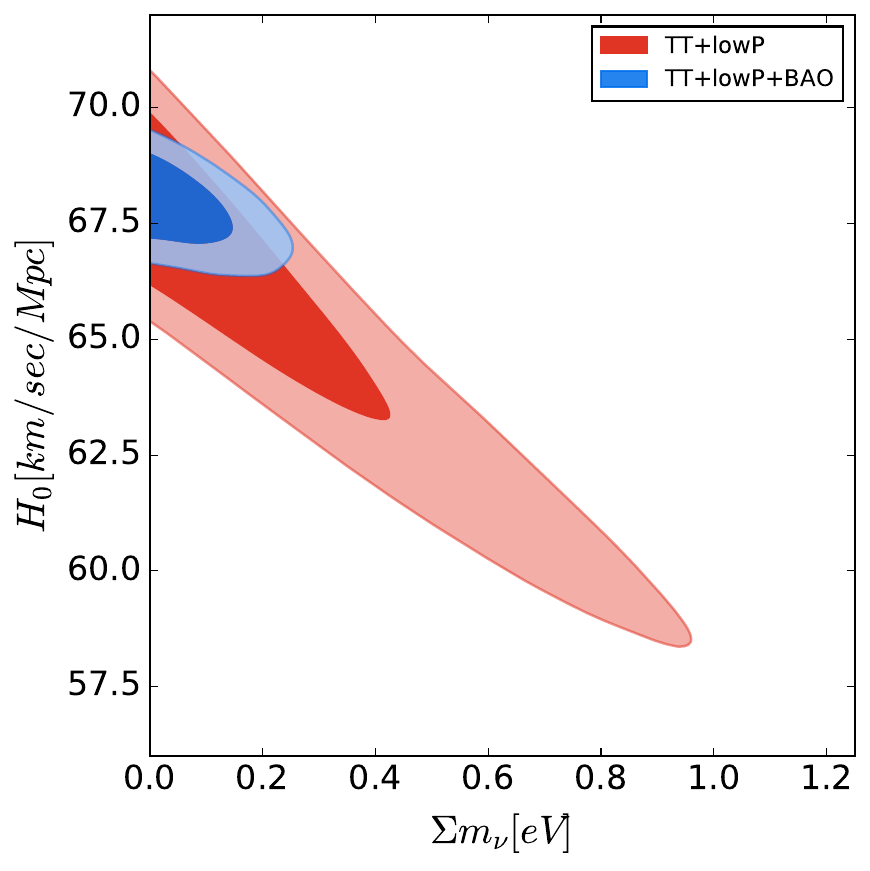}
\includegraphics[width=.4963\linewidth]{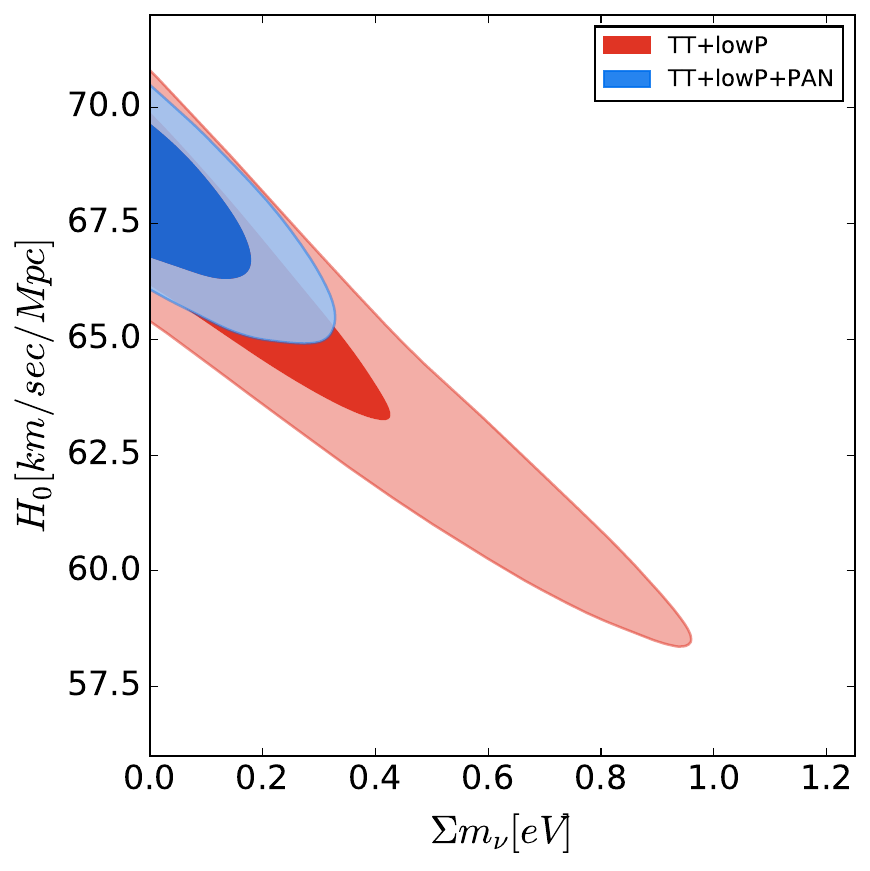}
\caption{\label{fig:6} 1$\sigma$ and 2$\sigma$ marginalised  contours for $H_0$ vs. $\sum m_{\nu}$ for TT+lowP, TT+lowP+BAO and TT+lowP+PAN datasets in $\Lambda CDM + \sum m_{\nu}$ model, showing the degeneracy breaking effect of BAO and PAN datasets separately. Evidently the BAO data is more effective in breaking the degeneracy between the two parameters. }
\end{figure}

As stated before, the Pantheon Sample (PAN) is the newest dataset available on Supernovae type Ia luminosity distance measurements, replacing its predecessor, the Joint Light-curve Analysis (JLA) sample. Observations of SNe Ia at a range of redshifts ($0.01<z<2.3$ for the Pantheon Sample) can be used to measure the evolution of luminosity distance as a function of redshift, and thereby determining the evolution of the scale factor \cite{Astier:2000as}. This information can be used to constraint cosmological parameters like dark energy equation of state $w$, and $\Omega_m$. The PAN dataset also provides substantially stronger mass bound when added to the CMB data, albeit not as strong as BAO data. In particular, TT+lowP+PAN gives a bound of $\sum m_{\nu} < 0.260$ eV, whereas for TTTEEE+lowP+PAN we get $\sum m_{\nu} < 0.230$eV. The 1$\sigma$ constraints on the Hubble constant are $H_0 = 67.43^{+1.16}_{-0.96}$ and $H_0 = 67.22^{+0.98}_{-0.70}$ km/sec/Mpc respectively. These are higher than that of CMB only data but lower than that of CMB+BAO data, which explains the weaker bounds from the Pantheon Sample compared to BAO. On the other hand, inclusion of both BAO and PAN data with CMB produces bounds slightly stronger than CMB+BAO. The bound with TT+lowP+BAO+PAN is $\sum m_{\nu} < 0.190$ eV whereas, for TTTEEE+lowP+BAO+PAN, it is $\sum m_{\nu} < 0.145$ eV, both of which far below the $\sum m_{\nu} < 0.23$ eV bound quoted in \cite{Planck2015}. Figure~\ref{fig:1} depicts this effect of addition of BAO, PAN and BAO+PAN to the Planck data. Also from Figure~\ref{fig:2} we see that the CMB+BAO+PAN combination prefers a slightly higher value of $H_0$ than CMB+BAO. The degeneracy breaking between $H_0$ and $\sum m_{\nu}$ due to BAO and PAN can be visualized in Figure~\ref{fig:6}.

\subsubsection{Results with $\tau$ and $H_0$ priors}
\label{sec:level4:1:2} 
\begin{figure}[tbp]
\centering 
\includegraphics[width=.4963\linewidth]{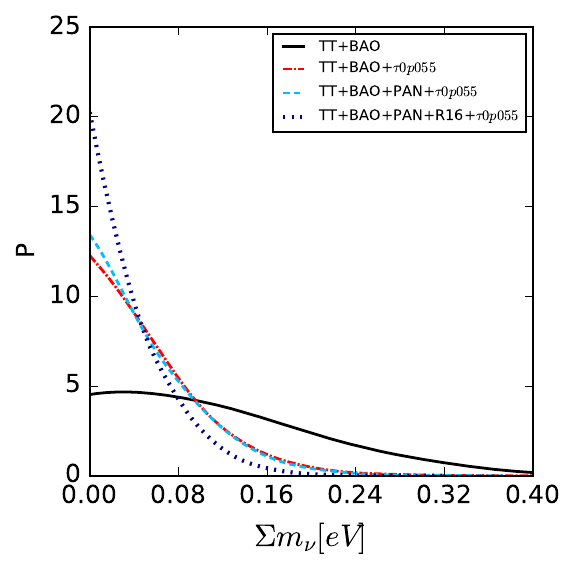}
\hfill
\includegraphics[width=.4963\linewidth]{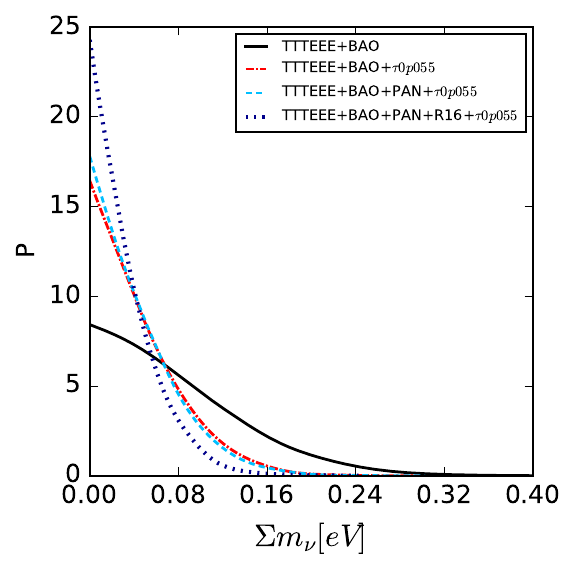}
\caption{\label{fig:4} Comparison of 1-D marginalized posterior distributions for $\sum m_{\nu}$ for various data combinations in $\Lambda CDM + \sum m_{\nu}$, with $\tau$ and $H_0$ priors. The plots are normalized in the sense that area under the curve is same for all curves. }
\end{figure}

\begin{figure}[tbp]
\centering 
\includegraphics[width=.4963\linewidth]{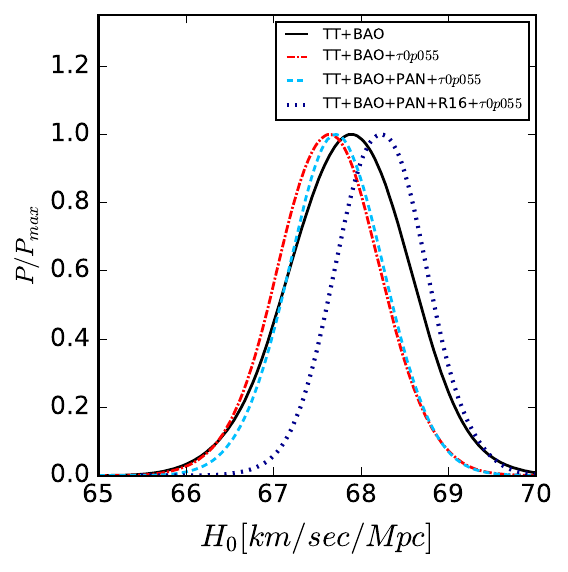}
\hfill
\includegraphics[width=.4963\linewidth]{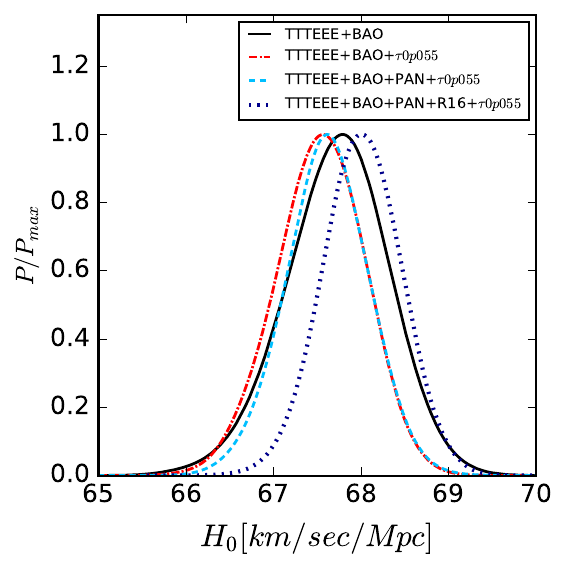}
\caption{\label{fig:5} Comparison of 1-D marginalized posterior distributions for $H_0$ for various data combinations in $\Lambda CDM + \sum m_{\nu}$, with $\tau$ and $H_0$ priors.}
\end{figure}

\begin{table}[tbp]
\centering
\begin{tabular}{@{}|l|r|@{}}
\hline
\multicolumn{2}{|c|}{Model: $\Lambda CDM + \sum m_{\nu}$} \\ 
\hline
Dataset                 & $\sum m_{\nu}$ (95\% C.L.)    \\ 
\hline
TT + BAO + $\tau 0p055$                  & < 0.159 eV                      \\
TT + BAO + FS + $\tau 0p055$             & < 0.159 eV                      \\
TT + BAO + PAN + $\tau 0p055$            & < 0.152 eV                      \\
TT + BAO - MGS + PAN + $\tau 0p055$      & < 0.141 eV                      \\
TT + BAO + FS + PAN + $\tau 0p055$       & < 0.160 eV                      \\
TT + BAO + SZ + $\tau 0p055$             & < 0.175 eV                      \\
TT + BAO + PAN + SZ + $\tau 0p055$       & < 0.168 eV                      \\
TT + lowP + R16                          & < 0.134 eV                       \\
TT + R16 + $\tau 0p055$                  & < 0.121 eV                      \\
TT + BAO + PAN + R16 + $\tau 0p055$      & < 0.117 eV                      \\
TT + BAO - MGS + PAN + R16 + $\tau 0p055$ & < 0.109 eV                      \\
TT + BAO + FS + PAN + R16 + $\tau 0p055$ & < 0.122 eV                      \\
\hline
\end{tabular}
\caption{\label{table:4} 95\% C.L. upper bounds on sum of three active neutrino masses in the degenerate case, in the backdrop of $\Lambda CDM + \sum m_{\nu}$ model for the given datasets.  Details about models and datasets are given in Section~\ref{sec:level2} and Section~\ref{sec:level3} respectively.}
\end{table}

\begin{table}[tbp]
\centering
\begin{tabular}{@{}|l|r|@{}}
\hline
\multicolumn{2}{|c|}{Model: $\Lambda CDM + \sum m_{\nu}$} \\ 
\hline
Dataset                 & $\sum m_{\nu}$ (95\% C.L.)    \\ 
\hline
TTTEEE + BAO + $\tau 0p055$                  & < 0.124 eV                      \\
TTTEEE + BAO + FS + $\tau 0p055$             & < 0.133 eV                      \\
TTTEEE + BAO + PAN + $\tau 0p055$            & < 0.118 eV                      \\
TTTEEE + BAO - MGS + PAN + $\tau 0p055$      & < 0.113 eV                      \\
TTTEEE + BAO + FS + PAN + $\tau 0p055$       & < 0.123 eV                      \\
TTTEEE + BAO + SZ + $\tau 0p055$             & < 0.136 eV                      \\
TTTEEE + BAO + PAN + SZ + $\tau 0p055$       & < 0.131 eV                      \\
TTTEEE + lowP + R16                          & < 0.125 eV                      \\   
TTTEEE + BAO + PAN + R16 + $\tau 0p055$      & < 0.091 eV                      \\
TTTEEE + BAO - MGS + PAN + R16 + $\tau 0p055$ & < 0.089 eV                      \\
TTTEEE + BAO + FS + PAN + R16 + $\tau 0p055$ & < 0.098 eV                      \\
\hline
\end{tabular}
\caption{\label{table:5} 95\% C.L. upper bounds on sum of three active neutrino masses in the degenerate case, in the backdrop of $\Lambda CDM + \sum m_{\nu}$ model for the given datasets. This is same as Table~\ref{table:4} but including the high-$l$ polarization data of Planck 2015. Details about models and datasets are given in Section~\ref{sec:level2} and Section~\ref{sec:level3} respectively.}
\end{table}

In the previous section (\ref{sec:level4:1:1}) we described how lower values of $\tau$ and higher values of $H_0$ help in constraining $\sum m_{\nu}$. Thus precise measurement of these two parameters are instrumental in obtaining meaningful bounds on the sum of neutrino masses. Figures~\ref{fig:4} and \ref{fig:5} shows 1-D marginalized posterior distributions for $\sum m_{\nu}$ and $H_0$ respectively, for various data combinations. In Tables~\ref{table:4} and \ref{table:5} we have presented the 95\% C.L. bounds on $\sum m_{\nu}$ where we have utilized the $\tau 0p055$ and R16 priors, along with bounds where we have included the FS and SZ datasets.

\begin{figure}[tbp]
\centering 
\includegraphics[width=.6\linewidth]{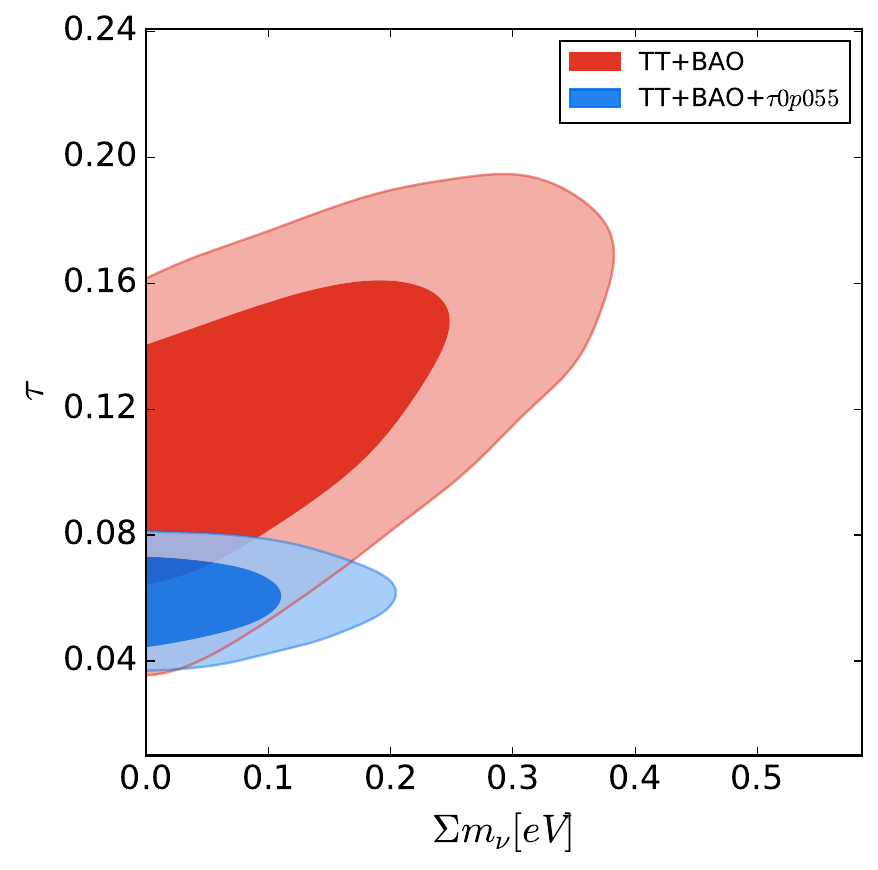}
\caption{\label{fig:9} 1$\sigma$ and 2$\sigma$ marginalized contours for $\tau$ vs. $\sum m_{\nu}$ for TT+BAO and TT+BAO+$\tau 0p055$ datasets in $\Lambda CDM + \sum m_{\nu}$ model, showing the reduction in correlation between $\tau$ and $\sum m_{\nu}$ due to addition of $\tau 0p055$, leading to a stronger bound on $\sum m_{\nu}$.}
\end{figure}

The addition of the Gaussian prior $\tau = 0.055 \pm 0.009$ significantly improves the bound by strongly breaking the degeneracy between $\tau$ and $\sum m_{\nu}$, which is depicted in Figure~\ref{fig:9} Compared to the bound of $\sum m_{\nu} < 0.311$ eV from TT+BAO, TT+BAO+$\tau 0p055$ yields a bound of $\sum m_{\nu} < 0.159$ eV. This change in mass bound can be attributed to a large change in the preferred value of $\tau$, mostly driven by the prior on $\tau$ (and albeit preferring a slightly lower value of $H_0$ as depicted in Figure~\ref{fig:5}). For TT+BAO we have the 1$\sigma$ bound of $\tau = 0.123\pm 0.031$, whereas for TT+BAO+$\tau 0p055$ we have $\tau = 0.060\pm 0.009$. Similarly for TTTEEE+BAO, we have  $\sum m_{\nu} < 0.204$ eV and $\tau = 0.105 \pm 0.023$, and it improves to $\sum m_{\nu} < 0.124$ eV and $\tau = 0.060^{+0.08}_{-0.09}$ for TTTEEE+BAO+$\tau 0p055$. We emphasize here again that this use of the prior $\tau 0p055$ is well motivated in the sense that, as Planck Collaboration \cite{Planck2016-intermediate} has mentioned in their paper, (1) it is the most accurate bound we currently have on $\tau$; (2) such a small value of $\tau$ also fully agrees with other astrophysical measurements of reionization from high redshift sources. For $\Lambda CDM + \sum m_{\nu}$, our tightest bound (except when we remove the MGS data from BAO) without any $H_0$ prior comes from addition of the PAN data. TTTEEE+BAO+PAN+$\tau 0p055$ gives a bound of $\sum m_{\nu} < 0.118$ eV, whereas without the high-$l$ polarization data, we achieved $\sum m_{\nu} < 0.152$ eV.  This is one of our main results in this paper, and one of the strongest bounds in literature available presently without the use of any $H_0$ prior.

\begin{subequations}\label{eq:4.2}
\begin{align}
\label{eq:4.2:1}
\sum m_{\nu} & < 0.152 ~\textrm{eV} ~(95\%) ~(\textrm{TT+BAO}+\textrm{PAN}+\tau 0p055), \\
\label{eq:4.2:2}
\sum m_{\nu} & < 0.118 ~\textrm{eV} ~(95\%) ~(\textrm{TTTEEE+BAO}+\textrm{PAN}+\tau 0p055).
\end{align}
\end{subequations}

\begin{figure}[tbp]
\centering 
\includegraphics[width=.6\linewidth]{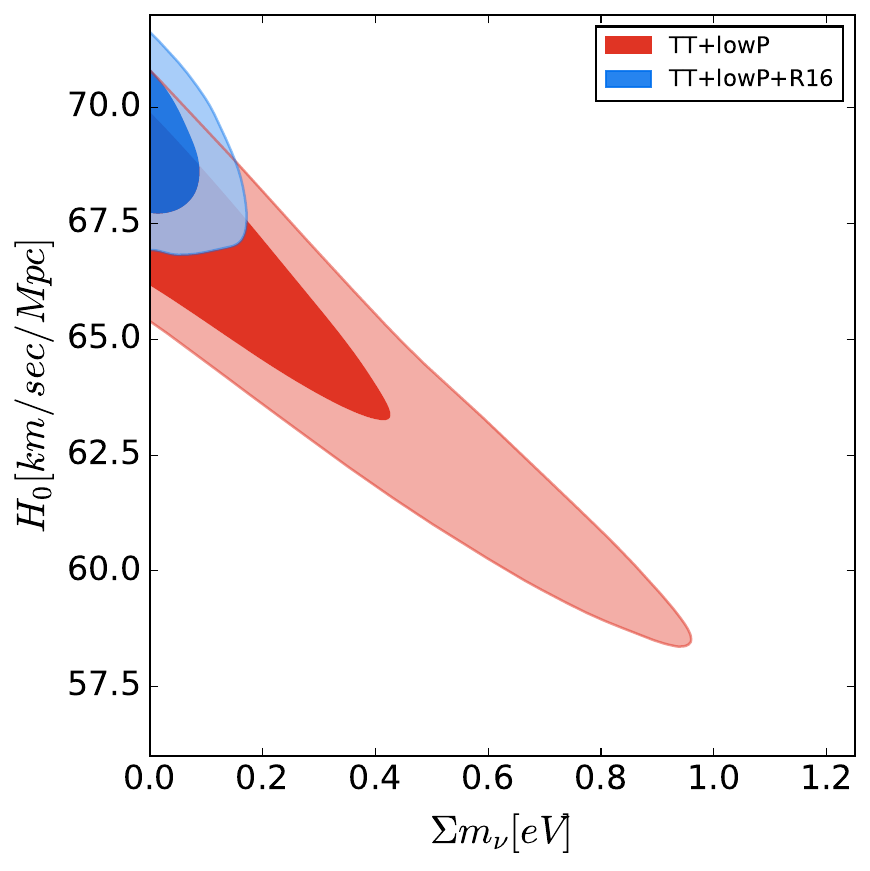}
\caption{\label{fig:7} 1$\sigma$ and 2$\sigma$ marginalized contours for $H_0$ vs. $\sum m_{\nu}$ for TT+lowP and TT+lowP+R16 datasets in $\Lambda CDM + \sum m_{\nu}$ model, showing the reduction in correlation between $H_0$ and $\sum m_{\nu}$ due to addition of the R16 prior, leading to a very strong bound on $\sum m_{\nu}$.}
\end{figure}

A prior on $H_0$ helps to break the degeneracy between $\sum m_{\nu}$ and $H_0$ in the Planck data. In Figure~\ref{fig:7} we demonstrate the same. Addition of the R16 prior ($H_0 = 73.24\pm 1.74$ km/sec/Mpc) leads to even stronger bounds than BAO data; TT+lowP+R16 yields $\sum m_{\nu} <$ 0.134 eV at 95\% C.L., whereas with TTTEEE+lowP+R16 it is $\sum m_{\nu} <$ 0.125 eV. A very aggressive bound of $\sum m_{\nu} < 0.091$ eV for $\Lambda CDM + \sum m_{\nu}$ is obtained with TTTEEE+BAO+PAN+R16+$\tau 0p055$, while the bound with TT+BAO+PAN+R16+$\tau 0p055$ is a bit relaxed at $\sum m_{\nu} < 0.117$ eV. These might be the most stringent bounds ever reported in literature within the minimal $\Lambda CDM + \sum m_{\nu}$ model. However, note that in Table~\ref{table:4}, TT + R16 + $\tau 0p055$  yields a bound of $\sum m_{\nu} <$ 0.121 eV, which shows us that BAO and PAN do not contribute significantly above the combination of CMB+R16. One can visualize from Figure~\ref{fig:4} that the R16 data prefers neutrinos with lower mass much more, due to the preference of significantly higher values of $H_0$ as shown in Figure~\ref{fig:5} and the strong anti-correlation present between $H_0$ and $\sum m_{\nu}$. However, as stated before, we need to be cautious with the interpretation of such tight mass bounds, since they are driven by the large 3.4 $\sigma$ tension between Planck and R16 measurements of the Hubble constant and since there seems to be no agreement among datasets on the value of $H_0$.  
While we do not use the lensing data, bounds with Planck 2015 and 2018 lensing data can be found in \cite{Planck2015} and \cite{Aghanim:2018eyx} respectively.

We notice the bounds can be strengthened further by removal of the DR7 Main Galaxy Sample (MGS) from the BAO data, as can be seen in Tables~\ref{table:4} and \ref{table:5}. We have denoted the MGS removed dataset simply as BAO -- MGS. We find that TT + BAO -- MGS + PAN + $\tau 0p055$ prefers an $H_0 = 67.88^{+0.55}_{-0.56}$ km/sec/Mpc which is a bit higher than TT + BAO + PAN + $\tau 0p055$, which prefers $H_0 = 67.71 \pm 0.55$ km/sec/Mpc. The preference of MGS sample for lower $H_0$ values has been discussed in \cite{doi:10.1093/mnras/stv154}. The lack of MGS data improves the mass bounds to $\sum m_{\nu} < 0.141$ eV for TT + BAO -- MGS + PAN + $\tau 0p055$, and $\sum m_{\nu} < 0.113$ eV for TTTEEE + BAO -- MGS + PAN + $\tau 0p055$. Adding the R16 prior, we get $\sum m_{\nu} < 0.109$ eV and $\sum m_{\nu} < 0.089$ eV respectively.  

\begin{figure}[tbp]
\centering 
\includegraphics[width=.6\linewidth]{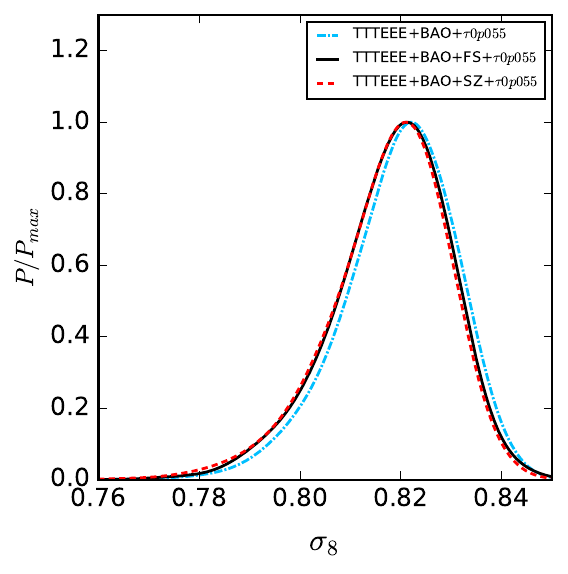}
\caption{\label{fig:8}  Comparison of 1-D marginalized posterior distributions for $\sigma_8$ for various data combinations in $\Lambda CDM + \sum m_{\nu}$ model. Data combinations with FS and SZ prefer a slightly lower value of $\sigma_8$, due to which slightly less stringent upper bounds on $\sum m_{\nu}$ are obtained.}
\end{figure}

Inclusion of the galaxy cluster data from full spectrum measurements (FS) from the SDSS-III BOSS DR12 either worsened or did not help the bounds, as can be seen in Tables~\ref{table:4} and \ref{table:5}. Previous studies  \cite{PhysRevD.96.123503,Zhao:2016ecj,Hamann:2010pw} have shown that the constraining power of the BAO measurements is higher than that of the full shape measurements in the minimal $\Lambda CDM + \sum m_{\nu}$ model, and we find that still to be true for the latest data. Addition of the galaxy cluster data (SZ) from the SPT-SZ survey also worsened the neutrino mass bounds slightly. As shown in Figure~\ref{fig:8}, both FS and SZ data prefer a slightly lower value of $\sigma_8$ (the normalization of matter power spectrum on scales of $8h^{-1}$ Mpc) and thereby favouring slightly larger values of $\sum m_{\nu}$; as more suppression of matter power spectrum allows for a larger neutrino mass sum, i.e., $\sigma_8$ and $\sum m_{\nu}$ are strongly anti-correlated. 

\subsection{Results for the $\Lambda CDM+r+\sum m_{\nu}$ Model}
\label{sec:level4:2}
\begin{table}[]
\centering
\begin{tabular}{@{}|l|r|@{}}
\hline

\multicolumn{2}{|c|}{Model: $\Lambda CDM + r + \sum m_{\nu}$} \\ \hline
Dataset                 & $\sum m_{\nu}$ (95\% C.L.)    \\ \hline
TT + BAO + PAN + $\tau 0p055$                             & < 0.161 eV                      \\
TT + BAO + PAN + BK14 + $\tau 0p055$                      & < 0.133 eV                      \\
TT + BAO + PAN + BK14 + R16 + $\tau 0p055$                & < 0.107 eV                      \\
\hline
TTTEEE + BAO + PAN + $\tau 0p055$                         & < 0.122 eV                      \\
TTTEEE + BAO + PAN + BK14 + $\tau 0p055$                  & < 0.110 eV                      \\
TTTEEE + BAO + PAN + BK14 + R16 + $\tau 0p055$            & < 0.085 eV                      \\ 
\hline
\end{tabular}
\caption{\label{table:8} 95\% C.L. upper bounds on sum of three active neutrino masses in the degenerate case, in the backdrop of $\Lambda CDM + r + \sum m_{\nu}$ model for the given datasets. Details about models and datasets are given in Section~\ref{sec:level2} and Section~\ref{sec:level3} respectively.}
\end{table}

In this section we present results in the $\Lambda CDM + r + \sum m_{\nu}$ model in Table~\ref{table:8}. For the TT+BAO+PAN+$\tau 0p055$ dataset, we see that in the $\Lambda CDM + r + \sum m_{\nu}$ model $\sum m_{\nu} <$ 0.161 eV, which is a bit relaxed than the $\sum m_{\nu} <$ 0.152 eV in the minimal $\Lambda CDM + \sum m_{\nu}$ model. This is simply due to added degeneracies in a extended parameter space with an extra parameter, $r$, which is the tensor-to-scalar ratio defined at a pivot scale of $k=0.005$ Mpc$^{-1}$. The TT+BAO+PAN+$\tau 0p055$ combination constrains the tensor-to-scalar ratio at $r <$0.13 (95\% C.L.), whereas for TTTEEE+BAO+PAN+$\tau 0p055$, we have $r <$ 0.12 (95\% C.L.). Addition of the BK14 data from BICEP/Keck collaboration, which contains information about the CMB BB spectra, strengthens this bound to $r <$ 0.07 for both the data combinations. It also strengthens the sum of neutrino mass bounds to $\sum m_{\nu} <$ 0.133 eV and $\sum m_{\nu} <$ 0.110 eV for TT+BAO+PAN+BK14+$\tau 0p055$ and TTTEEE+BAO+PAN+BK14+$\tau 0p055$ respectively, which are actually lower than the ones quoted in Eq.~\ref{eq:4.2}. 

CMB B-mode polarization has two well-known sources \cite{Ade:2014xna}. The first part comes from the inflationary gravitational waves (IGW), i.e., tensors, which is supposed to produce a bump peaked around $l\simeq 80$ (the so called 'recombination bump') in the BB-mode CMB spectra due to induction of quadruple anisotropies in the CMB within the last scattering surface. The IGW signature cannot be reproduced by scalar perturbations, and the amplitude of the bump depends on the tensor-to-scalar ratio, $r$. The other part comes from the deflection of CMB photons due to gravitational lensing produced by large scale structure at considerably late times, which converts a small fraction of the E mode power into B mode. This lensing BB spectra peaks at around $l\simeq 1000$. 
The 'reionization bump' is also expected to be present as in the EE spectra, in the $l <$ 10 region. However, the BK14 data contains information only in the $20<l<330$ and cannot constrain $\tau$ through the reionization bump.   

\begin{figure}[tbp]
\centering 
\includegraphics[width=.4963\linewidth]{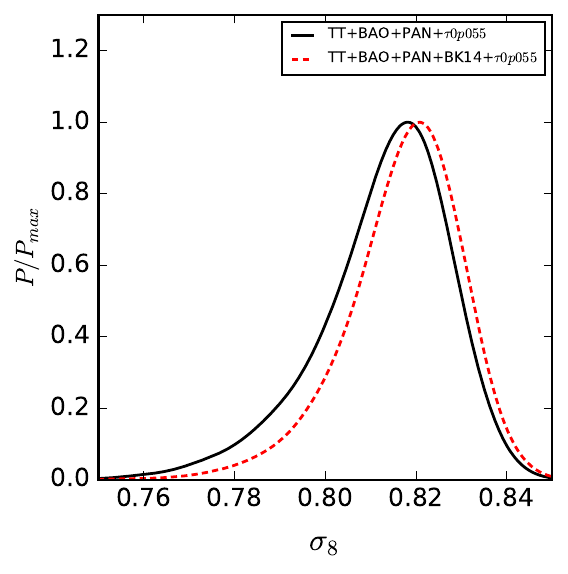}
\includegraphics[width=.4963\linewidth]{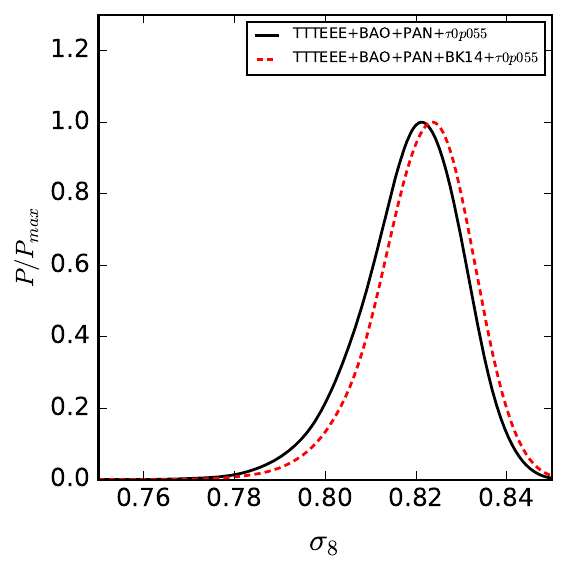}
\caption{\label{fig:10}  Comparison of 1-D marginalized posterior distributions for $\sigma_8$ for various data combinations in $\Lambda CDM + r + \sum m_{\nu}$ model. Addition of BK14 data seems to prefer a higher $\sigma_8$, due to which slightly more stringent upper bounds on $\sum m_{\nu}$ are obtained.}
\end{figure}

While the bound on $r$ is stronger due to BK14, this does not seems to be the main effect in tightening of the mass bounds.  We found that the correlation coefficient (defined as $R_{ij} = C_{ij}/\sqrt{C_{ii} C_{jj}}$, where $i$ and $j$ are the two parameters being considered and $C$ is the covariance matrix of cosmological parameters) between $r$ and $\sum m_{\nu}$ to be $R_{r,\Sigma m_{\nu}} = +0.056$ in case of TT+BAO+PAN+$\tau 0p055$, and $R_{r,\Sigma m_{\nu}} = +0.051$ in case of TT+BAO+PAN+BK14+$\tau 0p055$, which implies that the correlation is very small before addition of BK14, and there is also no big enough change in the correlation with the addition of BK14 dataset to account for the change in mass bound. The main effect might be coming from lensing BB spectra. Quantitatively, the correlation coefficient between $\sigma_8$ and $\sum m_{\nu}$ in TT+BAO+PAN+$\tau 0p055$ is $R_{\sigma_8,\Sigma m_{\nu}}=-0.828$, and in TT+BAO+PAN+BK14+$\tau 0p055$ it is $R_{\sigma_8,\Sigma m_{\nu}}=-0.780$. We find that BK14 data prefers a slightly larger value of $\sigma_8$ (see in Figure~\ref{fig:10}), and due to the strong anti-correlation present between  $\sigma_8$ and $\sum m_{\nu}$ in the data, the mass bounds improve a bit.  Similar inference can be made for the results including the high-$l$ polarization from Planck. As before, inclusion of the R16 prior improves the bounds even more. For TT+BAO+PAN+BK14+R16+$\tau 0p055$, we have a bound of $\sum m_{\nu} <$ 0.107 eV and for TTTEEE+BAO+PAN+BK14+R16+$\tau 0p055$ it is $\sum m_{\nu} <$ 0.085 eV, both of which are tighter than the corresponding bounds in minimal $\Lambda CDM + \sum m_{\nu}$ model without the BK14 data.

\subsection{Results for the $w_0 w_a CDM + \sum m_{\nu}$ Model (DDE)}
\label{sec:level4:3}
\begin{table}[tbp]
\centering
\begin{tabular}{@{}|l|r|@{}}
\hline
\multicolumn{2}{|c|}{Model: $w_0 w_a CDM + \sum m_{\nu}$ (DDE)}\\ \hline
Dataset                 & $\sum m_{\nu}$ (95\% C.L.)    \\  
\hline
TT + BAO + PAN + $\tau 0p055$                      & < 0.305 eV                      \\
TT + BAO + PAN + R16 + $\tau 0p055$                & < 0.284 eV                      \\
\hline
TTTEEE + BAO + PAN + $\tau 0p055$                  & < 0.276 eV                      \\
TTTEEE + BAO + PAN + R16 + $\tau 0p055$            & < 0.247 eV                      \\
\hline
\end{tabular}
\caption{\label{table:10} 95\% C.L. upper bounds on sum of three active neutrino masses in the degenerate case, in the backdrop of $w_0 w_a CDM + \sum m_{\nu}$ model (DDE), for the given datasets. Details about models and datasets are given in Section~\ref{sec:level2} and Section~\ref{sec:level3} respectively.}
\end{table}

\begin{figure}[tbp]
\centering 
\includegraphics[width=.4963\linewidth]{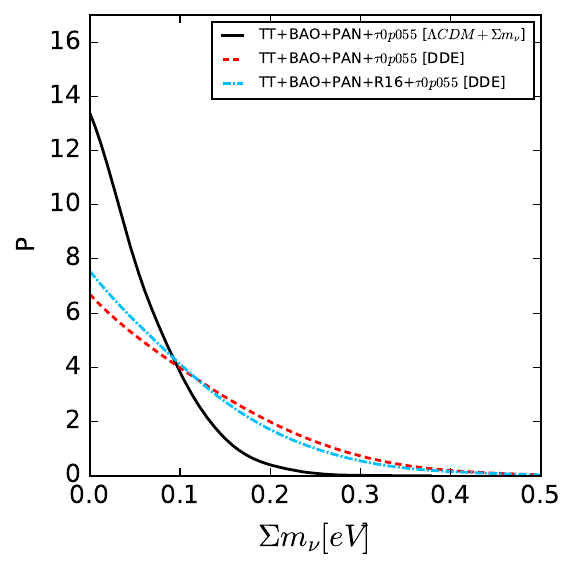}
\includegraphics[width=.4963\linewidth]{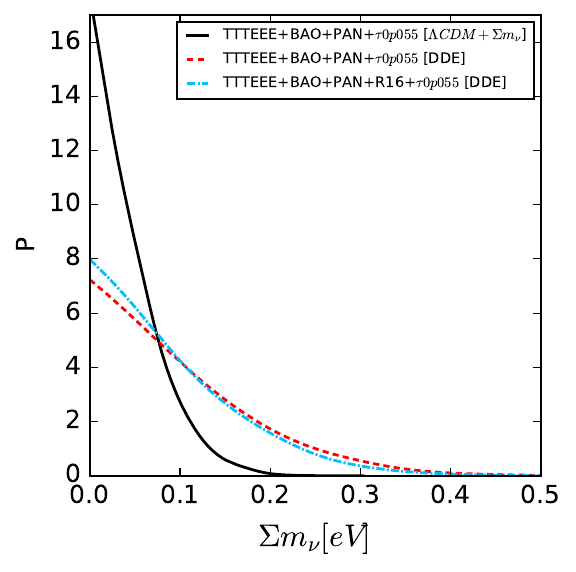}
\caption{\label{fig:11} Comparison of 1-D marginalised posterior distributions for $\sum m_{\nu}$  comparing the $\Lambda CDM + \sum m_{\nu}$ and DDE models. The plots are normalized in the sense that area under the curve is same for all curves. }
\end{figure}

\begin{figure}[tbp]
\centering 
\includegraphics[width=.4963\linewidth]{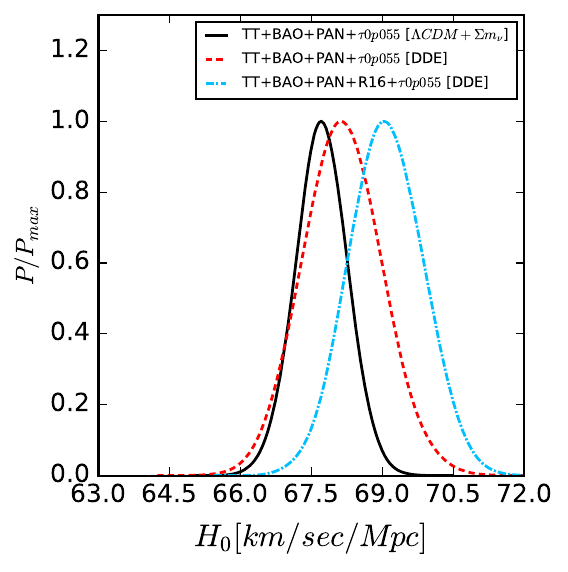}
\includegraphics[width=.4963\linewidth]{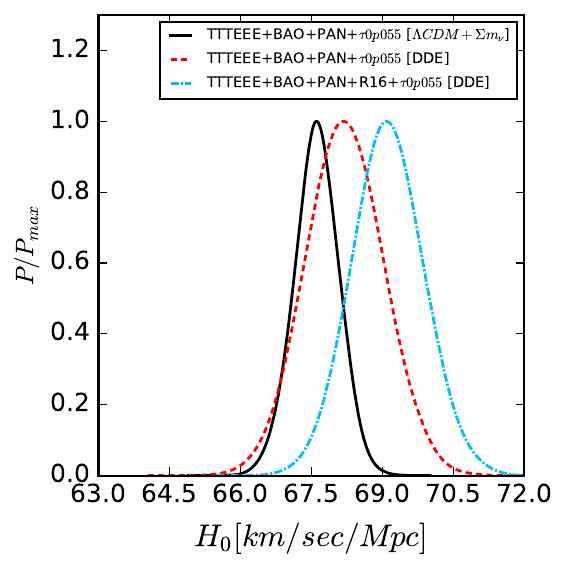}
\caption{\label{fig:12}  Comparison of 1-D marginalised posterior distributions for $H_0$ comparing the $\Lambda CDM + \sum m_{\nu}$ and DDE models. The DDE model prefers a broader distribution for $H_0$ and also the mean value of $H_0$ is higher, thereby reducing the tension between Planck 2015 and R16. Adding the R16 prior in the DDE model leads to even larger $H_0$ values.}
\end{figure}

\begin{figure}[tbp]
\centering 
\includegraphics[width=.6\linewidth]{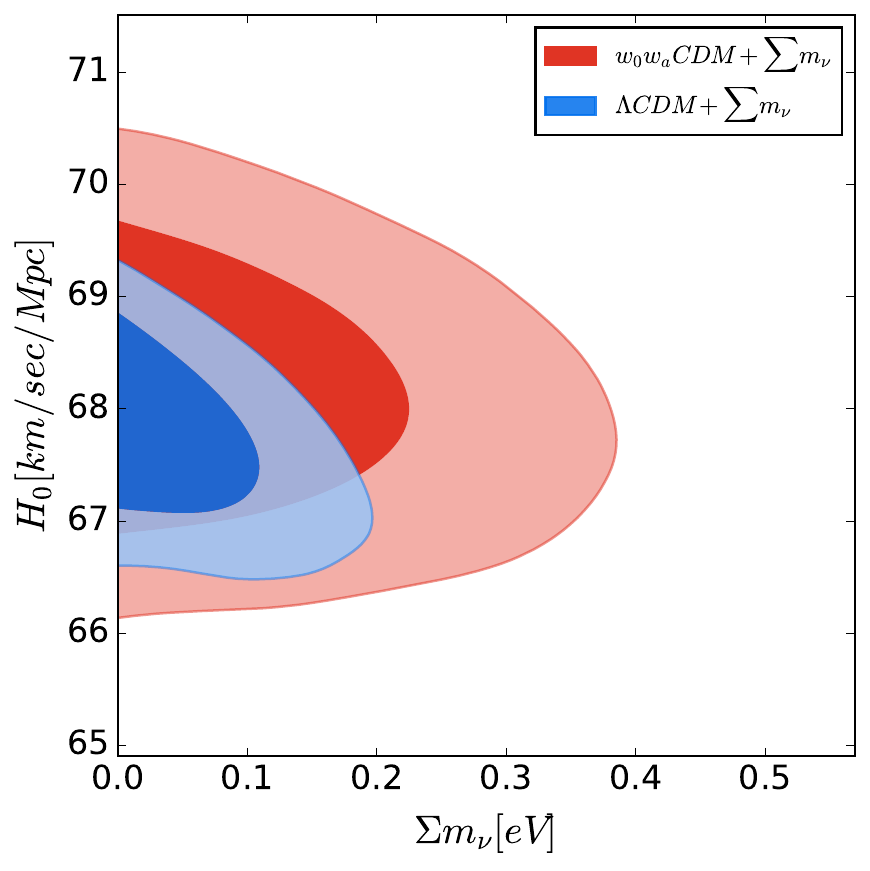}
\caption{\label{fig:17} 1$\sigma$ and 2$\sigma$ marginalised  contours in the $\sum m_{\nu}$ -- $H_0$ plane for TT+BAO+PAN+$\tau 0p055$, comparing their correlation in the $w_0 w_a CDM + \sum m_{\nu}$ (DDE) and $\Lambda CDM + \sum m_{\nu}$ models.}
\end{figure}

In this section we present results for the $w_0 w_a CDM + \sum m_{\nu}$ (DDE) model. The mass bounds are presented in Table~\ref{table:10}. For the DDE model we let the dark energy parameters vary in both the phantom and non-phantom range. There is a well-known strong degeneracy between the dark energy equation of state, $w$ and sum of  neutrino masses, $\sum m_{\nu}$ \cite{Hannestad:2005gj}. An increase in $\sum m_{\nu}$ can be compensated by a decrease in $w$, due to the mutual degeneracy with $\Omega_m$. This degeneracy leads to a large degradation of the mass bounds, as can be seen from Table~\ref{table:10} and comparing with the results from the $\Lambda CDM + \sum m_{\nu}$ model for the same datasets (see Tables~\ref{table:4} and \ref{table:5}). Figures~\ref{fig:11} and \ref{fig:12} provide the 1-D marginalized posterior distributions for $\sum m_{\nu}$ and $H_0$ respectively. From Figure~\ref{fig:11} we can clearly observe that for the same dataset, the DDE model allows much larger values of $\sum m_{\nu}$ than $\Lambda CDM + \sum m_{\nu}$. For TT+BAO+PAN+$\tau 0p055$ we obtain a bound of $\sum m_{\nu} <$ 0.305 eV, whereas for TTTEEE+BAO+PAN+$\tau 0p055$ the bound is slightly tighter at $\sum m_{\nu} <$ 0.276 eV. The dynamical dark energy model also helps to reduce the tension between Planck 2015 and R16, by allowing higher values of $H_0$ along with a broader distribution. (see Figure~\ref{fig:12}). Imposition of the R16 prior improves the mass bounds. However, the magnitude of this effect is less than what we saw in $\Lambda CDM + \sum m_{\nu}$. This is because $H_0$ and $w$ are also degenerate, i.e., a change in $H_0$ can be compensated by a change in $w$ instead of $\sum m_{\nu}$. This decreases the magnitude of correlation coefficient between $H_0$ and $\sum m_{\nu}$. This phenomenon of changing correlation across these two models can be looked upon in Figure~\ref{fig:17}. Quantitatively, for TT+BAO+PAN+$\tau 0p055$, the correlation coefficient  between $H_0$ and $\sum m_{\nu}$ changes from $R_{H_0,\Sigma m_{\nu}} = - 0.40$ in $\Lambda CDM + \sum m_{\nu}$ to $R_{H_0,\Sigma m_{\nu}} = - 0.15$ in $w_0 w_a CDM + \sum m_{\nu}$. Also, the DDE model and R16 have a much smaller tension than $\Lambda CDM + \sum m_{\nu}$ and R16.  


\begin{figure}[tbp]
\centering 
\includegraphics[width=.4963\linewidth]{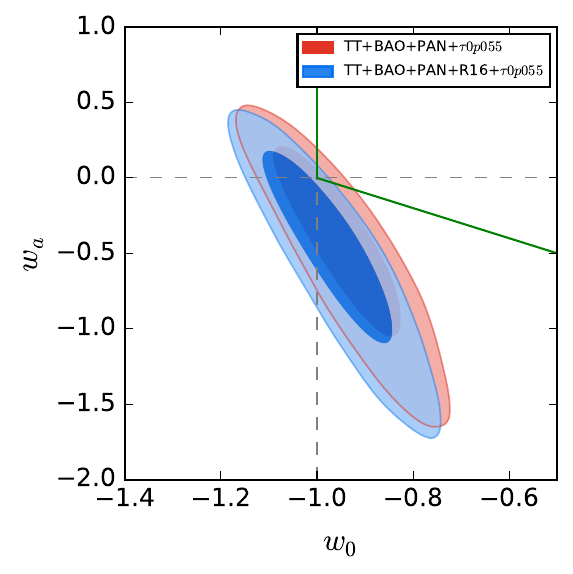}
\includegraphics[width=.4963\linewidth]{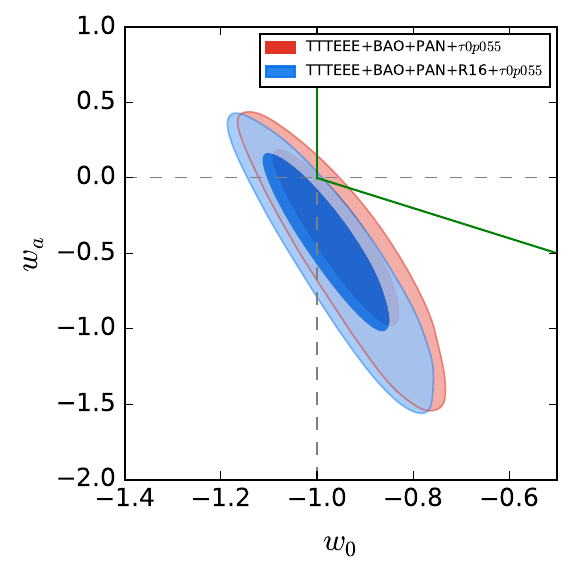}
\caption{\label{fig:16} 1$\sigma$ and 2$\sigma$ marginalized  contours for $w_0$ vs. $w_a$ for TT+BAO+PAN+$\tau 0p055$ and TTTEEE+BAO+PAN+R16+$\tau 0p055$ datasets in the $w_0 w_a CDM + \sum m_{\nu}$ (DDE) model. The dashed lines are at $w_0 = -1$ and $w_a = 0$ respectively. The two green lines originating from (-1,0) separate the non-phantom region from the rest. The region above the slanted green line and at the right of the vertical green line is the non-phantom region.}
\end{figure}

The $w_0$-$w_a$ diagram in Figure~\ref{fig:16} shows that for CMB+BAO+PAN+$\tau 0p055$ only a very small region which corresponds to completely non-phantom dark energy is allowed. Rest of the allowed region in the parameter space crosses the phantom barrier ($w=-1$ line) at some point in the evolution of the universe. We also find that the datasets are compatible with a cosmological constant ($w_0 = -1$, $w_a=0$). Imposing the R16 prior leads to shifting of the contours towards the phantom region. Thus, the allowed non-phantom region shrinks even more. A recent study \cite{DiValentino:2017zyq} showed that the disfavouring of the non-phantom region even persists in a 12 parameter extended space. Our results are also in agreement with Planck collaboration \cite{Planck2015} which reported similar contours for the given combination of similar but older datasets (see Figure 28 in that paper, for the combination of TT+lowP+ext, where 'ext' implies combination of BAO, JLA and a $H_0$ prior). In the next two sections we present our results on  neutrino mass bounds in a cosmology with only non-phantom dynamical dark energy. 

\emph{$\chi^2$-values}: 

Previous studies \cite{Zhao:2017cud,Feng:2017mfs} reported a improvement in fit with DDE models compared to $\Lambda \textrm{CDM}$. We found similar improvement in our analysis. We compare the best-fit $\chi^2$ values of the $w_0 w_a CDM + \sum m_{\nu}$ and $\Lambda CDM+\sum m_{\nu}$ models. We define, $\Delta \chi^2_{DDE} \equiv \chi^{2}_{\textrm{min}}(\textrm{DDE}) - \chi^{2}_{\textrm{min}}(\Lambda \textrm{CDM}+\sum m_{\nu})$, when used for the same dataset. For TT+BAO+PAN+$\tau 0p055$, we find  $\Delta \chi^2_{DDE} = - 0.40$; for  TTTEEE+BAO+PAN+$\tau 0p055$ it is $\Delta \chi^2_{DDE} = - 0.34$. The $\Delta \chi^2$ is better with the R16 prior. For TT+BAO+PAN+R16+$\tau 0p055$, we find $\Delta \chi^2_{DDE} = - 1.48$, whereas for TTTEEE+BAO+PAN+R16+$\tau 0p055$ it is  $\Delta \chi^2_{DDE} = - 3.27$. 

See also \cite{Yang:2017amu,Lambiase:2018ows,Poulin:2018zxs,Wang:2018ahw,Zhang:2017rbg,Zhang:2015uhk} for previous studies on massive neutrinos and dynamic dark energy together.

\subsection{Results for the $w_0 w_a CDM + \sum m_{\nu}$ Model with $w(z) \geq -1$ (NPDDE)}
\label{sec:level4:4}
\begin{table}[]
\centering
\begin{tabular}{@{}|l|r|@{}}
\hline
\multicolumn{2}{|c|}{Model: $w_0 w_a CDM + \sum m_{\nu}$ with $w(z) \geq -1$ (NPDDE)} \\ \hline
Dataset                 & $\sum m_{\nu}$ (95\% C.L.)    \\  \hline
TT + BAO + PAN + $\tau 0p055$                      & < 0.129 eV                      \\
TT + BAO + PAN + R16 + $\tau 0p055$                & < 0.106 eV                      \\
\hline
TTTEEE + BAO + PAN + $\tau 0p055$                  & < 0.101 eV                      \\
TTTEEE + BAO + PAN + R16 + $\tau 0p055$            & < 0.082 eV                      \\
\hline
\end{tabular}
\caption{\label{table:9} 95\% C.L. upper bounds on sum of three active neutrino masses in the degenerate case, in the backdrop of $w_0 w_a CDM + \sum m_{\nu}$ model with $w(z) \geq -1$ (NPDDE), for the given datasets. Details about models and datasets are given in Section~\ref{sec:level2} and Section~\ref{sec:level3} respectively.}
\end{table}

\begin{figure}[tbp]
\centering 
\includegraphics[width=.4963\linewidth]{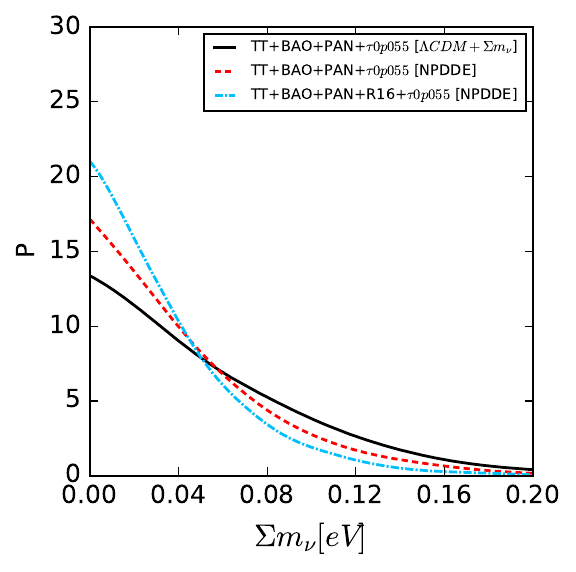}
\includegraphics[width=.4963\linewidth]{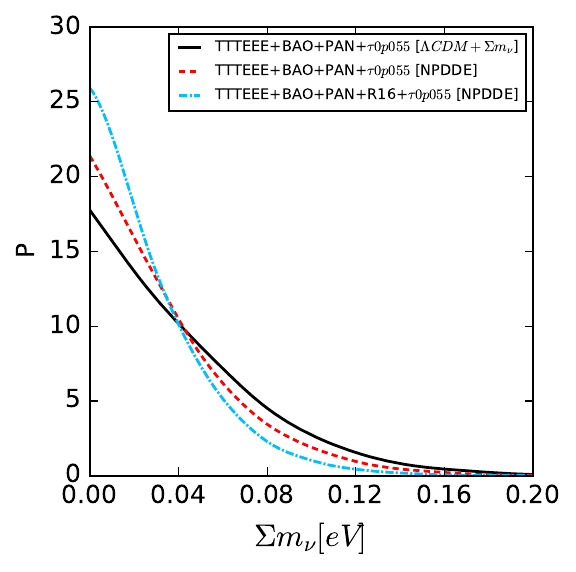}
\caption{\label{fig:13}  Comparison of 1-D marginalized posterior distributions for $\sum m_{\nu}$  comparing the $\Lambda CDM + \sum m_{\nu}$ and NPDDE models. The plots are normalized in the sense that area under the curve is same for all curves.}
\end{figure}
\begin{figure}[tbp]
\centering 
\includegraphics[width=.4963\linewidth]{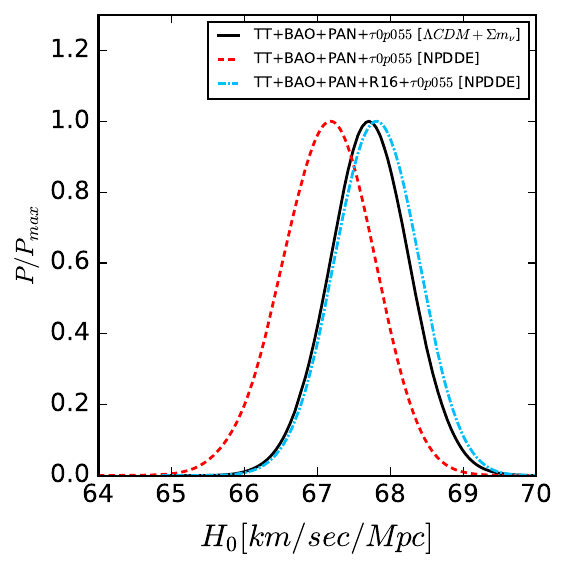}
\includegraphics[width=.4963\linewidth]{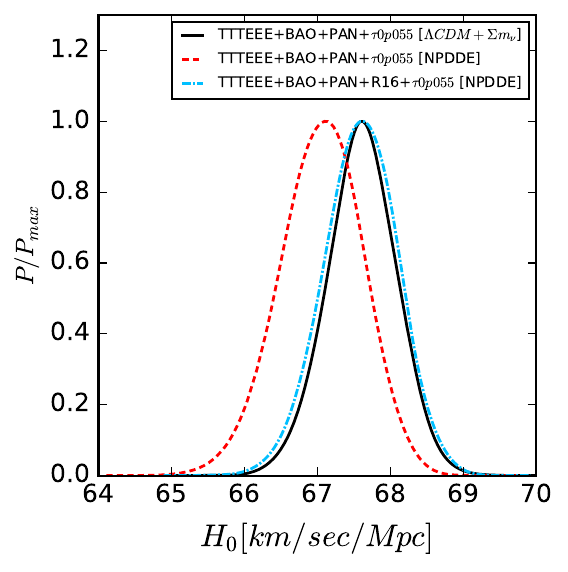}
\caption{\label{fig:14}  Comparison of 1-D marginalized posterior distributions for $H_0$ comparing the $\Lambda CDM + \sum m_{\nu}$ and NPDDE models. The NPDDE model prefers smaller values for $H_0$, thereby increasing the tension between Planck 2015 and R16. Adding the R16 prior in the NPDDE model leads to $H_0$ values which are somewhat similar to $\Lambda CDM + \sum m_{\nu}$ without R16.}
\end{figure}

While the current data prefers the phantom region  of the dark energy parameter space, it is also important to look at the non-phantom side of the things, since phantom dark energy is somewhat unphysical \cite{Vikman:2004dc}. Dark energy models with a single scalar field cannot cross the phantom barrier ($w=-1$) and more general models that permit the crossing require extra degrees of freedom to provide gravitational stability \cite{Fang:2008sn}. Field theories allowing phantom dark energy are fraught with one or more of the following problems like unstable vacuum, Lorentz violation, ghosts, superluminal modes, non-locality, or instability to quantum corrections.There, however, have also been theories where the field theory does not have any such issues but other effects like photon-axion conversion or modified gravity leads to an apparent $w<-1$ (see \cite{Ludwick:2017tox} for a brief review). Nonetheless, there are wide class of theories like quintessence \cite{Linder:2007wa,Caldwell:2005tm} which are non-phantom in nature and it is important to consider situations where we do not allow the phantom crossing.  

The constraints on $\sum m_{\nu}$ are shown in Table~\ref{table:9}. We find that the restricting ourselves to only the non-phantom sector yields bounds which are even stronger than the minimal $\Lambda CDM +\sum m_{\nu}$ model for the same datasets, even though it is an extended parameter space (also previously confirmed in \cite{Vagnozzi:2018jhn}). For TT + BAO + PAN + $\tau 0p055$, we have $\sum m_{\nu} < 0.129$ eV in the NPDDE model, whereas for $\Lambda CDM +\sum m_{\nu}$ model, using the same dataset, we had $\sum m_{\nu} < 0.152$ eV. For TTTEEE + BAO + PAN + $\tau 0p055$, in NPDDE, we have $\sum m_{\nu} < 0.101$ eV, compared to $\sum m_{\nu} < 0.118$ eV for $\Lambda CDM +\sum m_{\nu}$. Adding the R16 prior further reduces the allowable mass region, as we have seen throughout this paper. TT + BAO + R16 + PAN + $\tau 0p055$ prefers a $\sum m_{\nu} < 0.106$ eV, and TTTEEE + BAO + PAN + R16 + $\tau 0p055$ prefers $\sum m_{\nu} < 0.082$ eV, which is below the minimum sum required by the inverted hierarchy. 

However this substantial strengthening of neutrino mass bound in NPDDE model compared to DDE model is not surprising when we consider the degeneracy between $w$ and $\sum m_{\nu}$. As depicted in Figure 2 of \cite{Hannestad:2005gj}, due to strong anti-correlation between $w$ and $\sum m_{\nu}$, higher mass sum values prefer a lower value of $w$ and on the other hand higher values of $w$ for $w\geq -1$ are dominated by very low mass sum values. In NPDDE, what happens is we remove the phantom region, i.e., the portion of the parameter space which likes larger values of neutrino mass sum. Stronger bounds in an NPDDE model compared to $\Lambda CDM+\sum m_{\nu}$ is also confirmed in a recent study \cite{Vagnozzi:2018jhn}, which also confirmed the phenomenon that as we go away from the $w = -1$ line in the non-phantom region of the parameter space the mass bounds get stronger, whereas in the phantom region going away from the $w = -1$ line leads to weaker bounds, by running MCMC with separate fixed values of $w_0$ and $w_a$. A similar effect is seen in the bounds on $H_0$. Higher values of $H_0$ prefer a lower $w$, and removal of the phantom region of the parameter space leads to a preference towards lower values of $H_0$. Consequently, an NPDDE model actually increases the tension between Planck CMB data and R16. The alleviation of tension between Planck and R16 in DDE models comes from the phantom region of the $w_0-w_a$ plane. One of the consequences of such strong mass bounds is that, if in future neutrino hierarchy is found to be inverted by experiments, a universe with non-phantom dark energy will be less likely than a cosmological constant $\Lambda$ or phantom dark energy \cite{Vagnozzi:2018jhn}. The 1-D marginalized posteriors for $\sum m_{\nu}$ and $H_0$ for the NPDDE model are shown in Figures~\ref{fig:13} and \ref{fig:14} respectively.

\subsection{Results for the $w_0 w_a CDM + r + \sum m_{\nu}$ Model with $w(z) \geq -1$ (NPDDE+$r$)}
\label{sec:level4:5}
\begin{figure}[tbp]
\centering 
\includegraphics[width=.4963\linewidth]{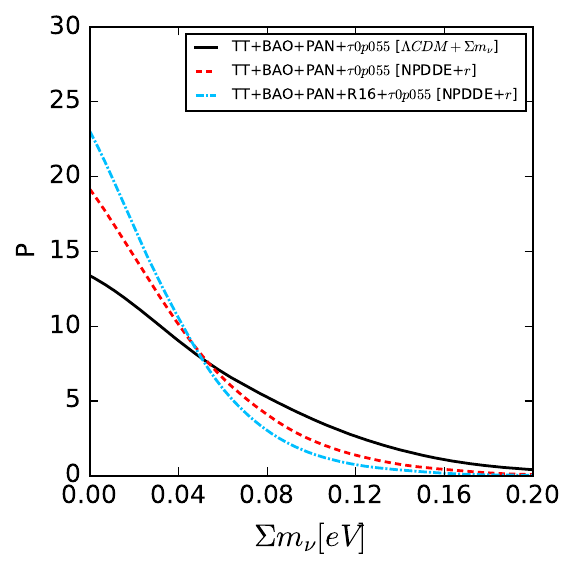}
\includegraphics[width=.4963\linewidth]{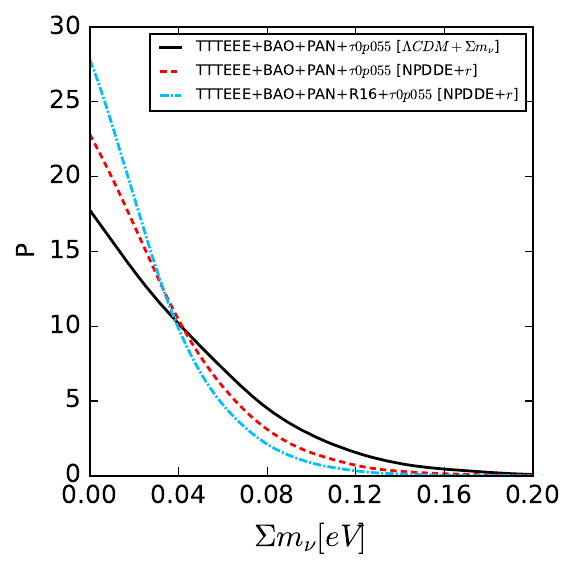}
\caption{\label{fig:18}  Comparison of 1-D marginalized posterior distributions for $\sum m_{\nu}$  comparing the $\Lambda CDM + \sum m_{\nu}$ and NPDDE+$r$ models. The plots are normalized in the sense that area under the curve is same for all curves. }
\end{figure}

\begin{figure}[tbp]
\centering 
\includegraphics[width=.4963\linewidth]{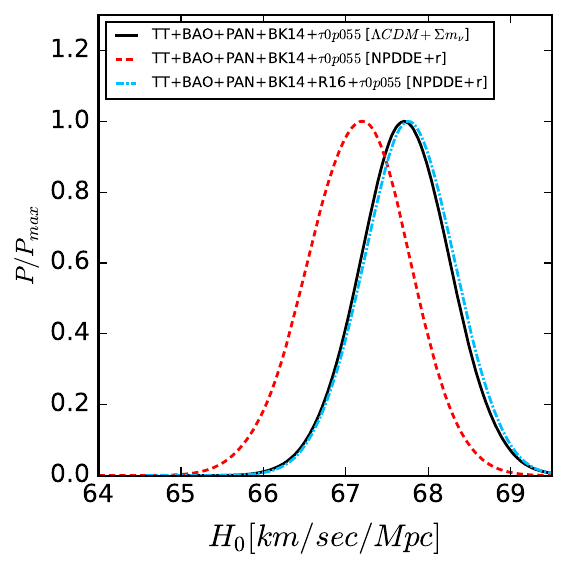}
\includegraphics[width=.4963\linewidth]{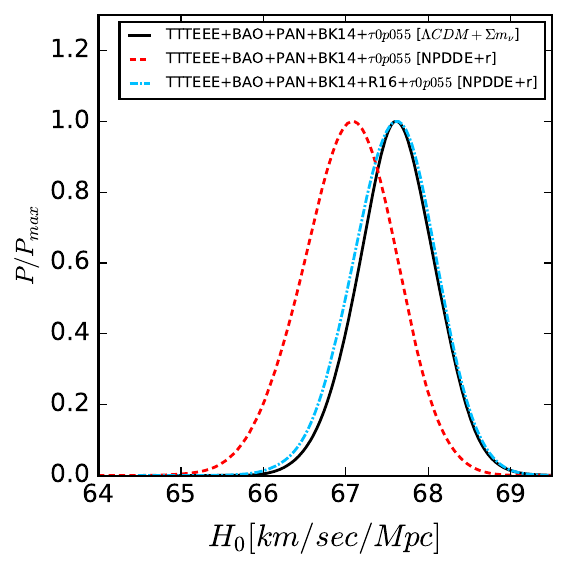}
\caption{\label{fig:19}  Comparison of 1-D marginalized posterior distributions for $H_0$ comparing the $\Lambda CDM + \sum m_{\nu}$ and NPDDE+$r$ models.}
\end{figure}

\begin{table}[tbp]
\centering
\begin{tabular}{@{}|l|r|@{}}
\hline
\multicolumn{2}{|c|}{Model: $w_0 w_a CDM + r + \sum m_{\nu}$ with $w(z) \geq -1$ (NPDDE+$r$)}\\ 
\hline
Dataset                 & $\sum m_{\nu}$ (95\% C.L.)    \\ 
\hline
TT + BAO + PAN + BK14 + $\tau 0p055$                      & < 0.116 eV                      \\
TT + BAO + PAN + BK14 + R16 + $\tau 0p055$                & < 0.095 eV                      \\
\hline
TTTEEE + BAO + PAN + BK14 + $\tau 0p055$                  & < 0.093 eV                      \\
TTTEEE + BAO + PAN + BK14 + R16 + $\tau 0p055$            & < 0.078 eV                      \\ 
\hline
\end{tabular}
\caption{\label{table:11} 95\% C.L. upper bounds on sum of three active neutrino masses in the degenerate case, in the backdrop of $w_0 w_a CDM + r +\sum m_{\nu}$ model with $w(z) \geq -1$ (NPDDE with tensors), for the given datasets. Details about models and datasets are given in Section~\ref{sec:level2} and Section~\ref{sec:level3} respectively.}
\end{table}

\begin{figure}[tbp]
\centering 
\includegraphics[width=.6\linewidth]{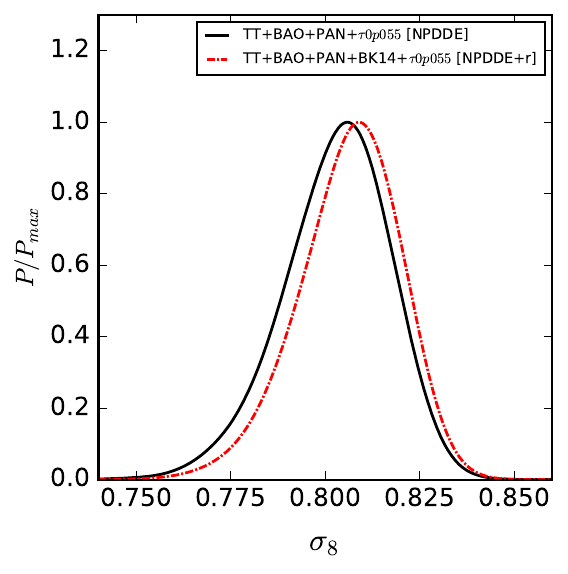}

\caption{\label{fig:15} Comparison of 1-D marginalised posterior distributions for $\sigma_8$ for NPDDE and NPDDE+$r$ models. Addition of BK14 data seems to prefer a higher $\sigma_8$, due to which slightly more stringent upper bound on $\sum m_{\nu}$ is obtained.}
\end{figure}

In this section we report results for the $w_0 w_a CDM + r + \sum m_{\nu}$ model with  $w(z) \geq -1$. We denote this model as "NPDDE+$r$". The main motivation behind studying this model was to see if we can further strengthen the mass bounds by adding the tensor-to-scalar ratio as a free parameter and adding the BK14 dataset, as in Section~\ref{sec:level4:2}. We find that it is still possible. Once again, the BK14 data prefers a slightly larger value of $\sigma_8$, as can be observed from Figure~\ref{fig:15}, which leads to slightly stronger bounds. The 1-D marginalized posterior distributions for $\sum m_{\nu}$ and $H_0$ are given in Figures~\ref{fig:18} and \ref{fig:19} respectively. The 95\% C.L. bounds on $\sum m_{\nu}$ are shown in Table~\ref{table:11}. Albeit the fact that we don't know for sure if we live in a universe with non-phantom dark energy or if the debatable R16 prior should be used, the $\sum m_{\nu} < 0.078$ eV bound for TTTEEE + BAO + PAN + BK14 + R16 + $\tau 0p055$ dataset for this NPDDE+$r$ model is possibly the strongest bound on $\sum m_{\nu}$ ever reported in literature for any kind of cosmological scenario.

\section{Discussion and Summary}
\label{sec:level5}
Neutrino oscillation experiments have confirmed that neutrinos are massive with three distinct species. However, still, certain neutrino properties including the sum of the three neutrino masses ($\sum m_{\nu}$) have not been precisely determined. Cosmology can put bounds on $\sum m_{\nu}$ and in reality, tightest bounds on $\sum m_{\nu}$ are obtained from cosmological data. Massive neutrinos leave distinct imprints in the CMB and can be constrained with CMB data. However since neutrinos with masses $\ll$ 1 eV are relativistic during decoupling of photons, CMB data is not particularly sensitive to low values of $\sum m_{\nu}$. Since massive neutrinos also cause suppression in the matter power spectrum, tighter bounds are obtained with large scale structure data. In this work we have used latest cosmological datasets available and provided very strong bounds on the sum of the masses of three active neutrinos in five different cosmological models: $\Lambda CDM + \sum m_{\nu}$, $\Lambda CDM+r+\sum m_{\nu}$, $w_0 w_a CDM + \sum m_{\nu}$ (DDE), $w_0 w_a CDM+\sum m_{\nu}$ with $w(z)\geq -1$ (NPDDE), and $w_0 w_a CDM+ r +\sum m_{\nu}$ with $w(z)\geq -1$ (NPDDE+$r$). Among datasets, along with CMB data from Planck 2015, we have used  BAO measurements from SDSS-III DR12, MGS and 6dFGS; SNe Ia luminosity distance measurements from Pantheon Sample (PAN); the BK14 data from the BICEP/Keck Collaboration; the galaxy cluster data from the SPT-SZ survey and suitable gaussian priors on $H_0$ (R16) and $\tau$ ($\tau 0p055$). The priors help in breaking the mutual degeneracies of $H_0$ and $\tau$ with $\sum m_{\nu}$ present in the Planck data. In the minimal $\Lambda CDM + \sum m_{\nu}$ model, we obtained a robust bound of $\sum m_{\nu} < 0.152$ eV at 95\% C.L. with the use of TT + BAO + PAN + $\tau 0p055$. Adding the high-$l$ polarization data tightens the bound to $\sum m_{\nu} < 0.118$ eV. The use of the $H_0$ prior further improves these bounds to $\sum m_{\nu} < 0.117$ eV and $\sum m_{\nu} < 0.091$ eV respectively, showing a weak preference for normal hierarchy. The low bounds obtained with the R16 prior, $H_0 = 73.24 \pm 1.74$ km/sec/Mpc are debatable since they are driven by the 3.4 $\sigma$ tension between Planck data and R16 over the value of $H_0$. Currently there seem to be no agreement over datasets on the value of $H_0$. The R16 prior itself is obtained from combining geometric distance calibrations of Cepheids, each of which separately give constraints on $H_0$: 72.25 $\pm$ 2.51, 72.04 $\pm$ 2.67, 76.18 $\pm$ 2.37, and 74.50 $\pm$ 3.27 km/sec/Mpc \cite{0004-637X-826-1-56}. Removing the third constraint (obtained from Milkyway cepheids) can reduce the $H_0$ tension and thereby worsen the bounds. While there is no reason to discard the data from Milkyway cepheids we should be cautious while looking at results obtained with the R16 prior. On the other hand, however, there is a possibility that both Planck and R16 might be correct and the discrepancy has to be explained by some new physics, like say, some dark radiation species which contributes to $N_\textrm{eff}$.

In the dynamical dark energy model $w_0 w_a CDM + \sum m_{\nu}$ (DDE) we find that the degeneracy between the dark energy equation of state, $w$ and $\sum m_{\nu}$ significantly relaxes the bounds. Our most conservative bound for this model is  
$\sum m_{\nu} < 0.305$ eV with TT + BAO + PAN + $\tau 0p055$, while the most aggressive bound of $\sum m_{\nu} < 0.247$ eV has been obtained with TTTEEE + BAO + PAN + R16 + $\tau 0p055$, which is very close to the  $\sum m_{\nu} < 0.23$ eV set by Planck collaboration in $\Lambda CDM + \sum m_{\nu}$ with similar datasets. This shows the superior constraining power of the new datasets and priors. The DDE model also provides marginally better $\chi^2$ fit to the data compared to $\Lambda CDM + \sum m_{\nu}$ and partially alleviates the $H_0$ tension between Planck data and R16. While we find that the DDE model is compatible with a cosmological constant for the combination of CMB+BAO+PAN+$\tau 0p055$, the 68\% and 95\% contours in the $w_0-w_a$ plane mostly allow phantom dark energy ($w < -1$) and only a very small region of non-phantom dark energy, which shrinks even more with the inclusion of the R16 prior. Also due to the strong degeneracy between $\sum m_{\nu}$ and $w$, larger $\sum m_{\nu}$ is preferred for lower $w$ (i.e. phantom region), while the deeper we go into the non-phantom region smaller the preferred $\sum m_{\nu}$. So in the NPDDE model ($w_0 w_a CDM+\sum m_{\nu}$ with $w(z)\geq -1$) when we vary the dark energy parameters only in the non-phantom region, we end up with $\sum m_{\nu}$ bounds which are even tighter than $\Lambda CDM + \sum m_{\nu}$ (also confirmed by a recent study \cite{Vagnozzi:2018jhn}). In NPDDE, without R16, we obtained a very strong bound of $\sum m_{\nu} < 0.101$ eV with TTTEEE + BAO + PAN + $\tau 0p055$. Adding the R16 prior leads to an even more aggressive bound of $\sum m_{\nu} < 0.082$ eV. Allowing for tensors in the $\Lambda CDM+r+\sum m_{\nu}$ model and including the BK14 data leads to slightly stronger bounds, which seems to be stemming from BK14 preferring a slightly larger value of $\sigma_8$. This phenomenon persists even when we consider the NPDDE model with tensors (i.e., NPDDE+$r$ model). In the NPDDE+$r$ model, without R16, for TTTEEE + BAO + PAN + BK14 + $\tau 0p055$, we found $\sum m_{\nu} <$ 0.093 eV. Such strong bounds in the NPDDE and NPDDE+r models imply that if future experiments discover that neutrino hierarchy is inverted, the nature of dark energy is more likely to be phantom than non-phantom ((as previously inferred in \cite{Vagnozzi:2018jhn}). In NPDDE+$r$, with the R16 prior, we find our most aggressive bound of $\sum m_{\nu} < 0.078$ eV. It might be the strongest bound ever quoted in any literature for any kind of cosmological model.

See \cite{ABAZAJIAN201566,Hall:2012kg,Wu:2014hta,Zhen:2015yba,Kitching:2008dp,Font-Ribera:2013rwa,Villaescusa-Navarro:2015cca,Mueller:2014dba,Errard:2015cxa} for forecasts on neutrino mass from possible future experiments. See also \cite{Gariazzo:2018pei,Capozzi:2018ubv, deSalas:2018bym} for the current status of determination of neutrino mass hierarchy from various experiments. We conclude with the remark that future experiments are expected to measure the optical depth to reionization very accurately and also reconstruct the CMB lensing potential accurately through precise measurements of CMB polarization. These will be instrumental in constraining $\sum m_{\nu}$ and there are exciting times ahead in neutrino cosmology.

\acknowledgments
SRC thanks the cluster computing facility at HRI (\url{http://www.hri.res.in/cluster/}). The authors would also like to thank the Department of Atomic Energy (DAE) Neutrino Project of HRI. This project has received funding from the European Union's Horizon 2020 research and innovation programme InvisiblesPlus RISE under the Marie Sklodowska-Curie grant agreement No 690575. This project has received funding from the European Union's Horizon 2020 research and innovation programme Elusives ITN under the Marie Sklodowska-Curie grant agreement No 674896. 

\bibliography{pantheon_numass}
\bibliographystyle{jhep}

\end{document}